\documentclass[10pt,journal,compsoc]{IEEEtran}

\usepackage{lipsum}

\usepackage{graphicx}
\graphicspath{{figures/}}
\DeclareGraphicsExtensions{.jpg,.png}
\usepackage{etoolbox}
\usepackage[caption=false,font=footnotesize]{subfig}

\usepackage{enumitem}
\usepackage{booktabs}

\usepackage{multirow,makecell}

\usepackage{framed}
\usepackage{float}

\usepackage{listings}
\usepackage{comment}
\usepackage{array}
\usepackage{rotating}
\usepackage{adjustbox}
\usepackage{tabularx}
\usepackage[framemethod=default]{mdframed}
\usepackage{tikz}
\usepackage{pgfplots, pgfplotstable}
\usetikzlibrary{patterns}

\usepackage[T1]{fontenc}
\usepackage{lmodern}

\usepackage{pgfplotstable}
\usepackage{filecontents}

\usepackage{array}
\usepackage{xcolor}
\usepackage{colortbl}
\PassOptionsToPackage{hyphens}{url}

\usepackage[hidelinks]{hyperref}
\usepackage{adjustbox}
\usepackage{fancyvrb}
\usepackage{balance}
\usepackage{xspace}
\newcommand{\etal}{\textit{et al.}\xspace}

\newcommand{\ie}{\textit{i.e.,}\xspace}
\newcommand{\eg}{\textit{e.g.,}\xspace}
\newcommand{\cf}{\textit{cf.}\xspace}

\newcounter{protocol}[section]

\newcolumntype{B}[2]{
    >{\adjustbox{angle=#1,lap=\width-(#2)}\bgroup}
    l
    <{\egroup}
}
\newcommand*\rotl{\multicolumn{1}{B{35}{1em}}}

\newcolumntype{Z}[2]{
    >{\adjustbox{angle=#1,lap=\width-(#2)}\bgroup}
    l
    <{\egroup}
}

\newcommand*\rotL{\multicolumn{1}{@{\hskip1pt}Z{90}{0.1em}}}
\newcommand*\rots{\multicolumn{1}{Z{90}{0em}}}
\newcommand*\rotR{\multicolumn{1}{Z{90}{1em}@{\hskip2pt}|}}

\newcolumntype{L}[1]{>{\raggedright\let\newline\\\arraybackslash\hspace{0pt}}m{#1}}
\newcolumntype{C}[1]{>{\centering\let\newline\\\arraybackslash\hspace{0pt}}m{#1}}
\newcolumntype{R}[1]{>{\raggedleft\let\newline\\\arraybackslash\hspace{0pt}}m{#1}}

\newcommand{\full}{\ $\bullet$}
\newcommand{\prt}{\ $\circ$}

\definecolor{semi-light-gray}{gray}{0.7}
\definecolor{light-gray}{gray}{0.8}

\newcommand{\uw}{\textsl{user-to-web}}
\newcommand{\ud}{\textsl{user-to-device}}
\newcommand{\dw}{\textsl{device-to-web}}

\usepackage{pgfplotstable} 

\ifCLASSOPTIONcompsoc
  \usepackage[nocompress]{cite}
\else
  \usepackage{cite}
\fi

\ifCLASSINFOpdf
\else
\fi

\begin{document}
\title{Comparative Analysis and Framework Evaluating Mimicry-Resistant and Invisible Web Authentication Schemes}

\author{
Furkan Alaca,
AbdelRahman Abdou,
Paul C. van Oorschot
\IEEEcompsocitemizethanks{\IEEEcompsocthanksitem F. Alaca is with the Department of Mathematical and Computational Sciences at the University of Toronto Mississauga, Canada. This work was done while he was a PhD candidate at the School of Computer Science at Carleton University.} \IEEEcompsocitemizethanks{\IEEEcompsocthanksitem A. Abdou is with the School of Computer Science, Carleton University, Canada.} \IEEEcompsocitemizethanks{\IEEEcompsocthanksitem P.C. van Oorschot is with the School of Computer Science, Carleton University, Canada.} \thanks{Version: \today.}
}

\maketitle

\begin{abstract}
Many password alternatives for web authentication proposed over the years, despite having different designs and objectives, all predominantly rely on the knowledge of some secret. This motivates us, herein, to provide the first detailed exploration of the integration of a fundamentally different element of defense into the design of web authentication schemes: a \emph{mimicry-resistance} dimension. We analyze web authentication mechanisms with respect to new \emph{usability} and \emph{security} properties related to mimicry-resistance (augmenting the UDS framework), and in particular evaluate \emph{invisible} techniques (those requiring neither user actions, nor awareness) that provide some mimicry-resistance (unlike those relying solely on static secrets), including device fingerprinting schemes, PUFs (physically unclonable functions), and a subset of Internet geolocation mechanisms.
\end{abstract}

\IEEEpeerreviewmaketitle

\section{Introduction}

None of the many schemes proposed over the years to replace password-based web authentication have offered sufficient usability and deployability benefits to displace passwords at the Internet scale~\cite{bonneau2012quest}. Since passwords do not seem to be disappearing~\cite{herley2012research}, a prominent avenue of improvement is to reinforce their security by parallel mechanisms~\cite{bonneau2015passwords} without further burdening users.

A challenge in providing sufficient security guarantees for web authentication, \ie \uw{} and \dw{} (versus \ud{}), is that the security of many schemes, including those relying on \emph{something-you-have} or \emph{something-you-are}, requires the ability to protect a secret. 
For example, a physical biometric such as a fingerprint can be captured in transit and replayed by an attacker without possession of the physical fingerprint, resulting in security properties similar to other stored secrets such as passwords. The reliance of many schemes on an element of secrecy is reflected in the Usability-Deployability-Security (UDS) evaluation framework~\cite{bonneau2012quest}, where eight of nine security properties assess a scheme's resilience against the exposure of a secret.

We revisit the process of compromising an account from an attacker's perspective, now viewing it as a two-stage process involving both exposure and mimicry. Exposure refers to the capture of information that enables account access, such as a password, session cookie, or a cryptographic key; mimicry refers to actions performed by the attacker to impersonate the legitimate user's behavior or associated characteristics, such as replaying a password or spoofing a user's geographic location (\ie in location-based authentication~\cite{denning1996location}). 
Accordingly, a scheme's authentication token may resist attacks by both: resisting exposure and resisting mimicry.

The term \emph{mimicry} has been used previously in the context of intrusion detection systems, referring to the attacker's ability to mimic legitimate traffic~\cite{wagner2002mimicry}. In contrast, the mimicry-resistance dimension has been relatively little-studied in the context of web authentication. We therefore investigate mimicry-resistance in web authentication herein and systematize elements not previously addressed in the literature. We first define a suitable set of criteria and then rank schemes across a continuum of three classes of resistance to mimicry, as detailed in Section~\ref{sec:classi}. Mimicry-resistance has been explored within the context of \ud{} authentication; for example, Khan et al.~\cite{khan2014comparative} analyze mimicry attacks on \ud{} authentication schemes that rely on behavioural biometrics, such as touchscreen finger movement patterns. In contrast, herein we analyze web authentication schemes; our analysis finds that most of these schemes offer little to no resistance to mimicry.

To construct a more comprehensive evaluation framework for authentication schemes, we augment the existing UDS security properties, which concentrate on exposure-resistance, with new properties measuring mimicry-resistance. 
We leverage the UDS framework security properties to systematically rate authentication schemes across an additional continuum ranging from lowest to highest resistance to exposure, and use these as orthogonal axes (exposure and mimicry) to plot a two-dimensional chart. Along both dimensions, our evaluation also reflects the scalability of attacks required to defeat a scheme.

The lack of mimicry-resistant schemes among those previously evaluated under UDS~\cite{bonneau2012quest} motivates us to select and evaluate a representative set of techniques for reinforcing web authentication that demonstrate the benefits of mimicry resistance. Our constructed framework dissects various degrees of mimicry resistance, and helps analyze currently known methodologies that may provide some degree of resistance to mimicry attacks. We then evaluate techniques that fall under four approaches, namely device fingerprinting (FP)~\cite{eckersley2010unique}, Internet geolocation~\cite{dong2012network}, Physically Unclonable Functions (PUFs)~\cite{yu2016pervasive}, and One Time Passwords (OTPs)~\cite{totp}; some variations of these offer resistance to mimicry attacks, and/or are also \emph{invisible}, in that they do not require any user effort to configure or use (see Section~\ref{sec:conveyor}). We supplement the UDS framework with two usability properties and four security properties (Table~\ref{tab:prims}). Under this revised framework, we evaluate techniques which fall under the four aforementioned approaches when combined with passwords, and find that invisible and mimicry-resistant schemes combined with passwords provide significantly higher resistance to attack. This constitutes an initial step towards identifying mimicry-resistant web authentication schemes that can enhance security with minimal usability penalties. 

In summary, the following contributions are made:
\begin{itemize}
\item Investigating the mimicry-resistance dimension in web authentication, including ranking schemes under three sub-classes of mimicry resistance.
\item Exploring newer invisible techniques, and evaluating their degree of mimicry resistance when used for web authentication.
\item Constructing a comprehensive evaluation framework, which includes (a) a two-dimensional chart combining the exposure and mimicry resistance dimensions, to visually reflect the ability of a scheme to resist scalable attacks; (b) an augmented UDS framework.
\item Using the augmented UDS framework herein for the first detailed exploration of the benefits of combining mimicry-resistant web authentication techniques with ubiquitous password-based \uw{} authentication.
\end{itemize}

The remainder of this paper is organized as follows. Section~\ref{sec:conveyor} provides background. Section~\ref{sec:classi} introduces the mimicry-resistance dimension in web authentication, and plots a representative subset of authentication schemes onto an Exposure-Mimicry two-dimensional space. Section~\ref{sec:added} evaluates relatively little-explored schemes found to have mimicry resistance, using a modified UDS framework with new properties addressing the mimicry resistance dimension. 
Section~\ref{sec:combined} analyzes benefits when the aforementioned schemes are combined with passwords. Section~\ref{sec:conclusion} concludes.

\section{Background and Context}
\label{sec:conveyor}

\begin{figure}
\centering
\includegraphics[scale=0.45]{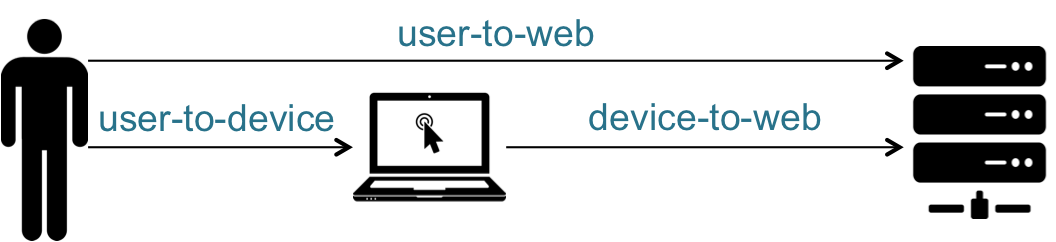}
\caption{A comparison between \ud{}, \dw{}, and \uw{} (directly).}
\label{fig:stickfigure}
\end{figure}

We briefly provide background and define terms for modes of authentication (Fig.~\ref{fig:stickfigure}) and review related work on the evaluation of authentication schemes.

The standard method of user authentication on the web is for the user to type a password into a web form, which is submitted over HTTP(S) for the web server to verify. We refer to this as \uw{} (see Fig.~\ref{fig:stickfigure}) authentication---even though the password is physically entered into the user's local device, the verifier is a remote web server. The device used for authentication, which we call the \emph{access device} herein, passively conveys the authentication token,\footnote{We use the terms \emph{token} (by default indicating a digital token such as a password) and \emph{credential} interchangeably. Hardware tokens will be explicitly identified.} and does no checking on behalf of the web server. In contrast, when a user authenticates to their smartphone, it is \ud{}. Some authentication schemes that appear to be \uw{} are actually \emph{two-stage} authentication schemes that combine a \ud{} scheme and a \dw{} scheme; for example, a mobile payment app may authenticate the user via a biometric (\eg fingerprint or iris scan), which unlocks a locally-stored cryptographic key used by the mobile device to authenticate to the remote server.

{\bf Implicit vs. invisible authentication. }
Implicit authentication~\cite{jakobsson2009implicit} schemes reduce user burden by authenticating users without requiring any deliberate user effort at login time by, \eg measuring and using users' biometric attributes or physiological behaviors for authentication. Many such schemes have been proposed~\cite{frank2013touchalytics}\cite{feng2014}\cite{khan2014comparative}, and are typically used for \ud{} authentication; the accuracy and resistance to mimicry of some of these schemes have been evaluated~\cite{khan2014comparative}. In contrast, herein we consider web authentication schemes, thus focusing on \uw{} and \dw{} authentication. As such, \ud{} (\eg~\cite{mare2014zebra}\cite{acar2018waca}) and similarly other paradigms such as \textsl{device-to-device} (\eg~\cite{aksu2018identification}\cite{radhakrishnan2015gtid}), are out of the scope of our work.

We also make the distinction between \emph{implicit} and \emph{invisible} schemes. The former refers to schemes that do not require extra user effort during login, but do require some initial user effort for setup; those are generally \ud{} schemes that authenticate the user based on their behaviour as measured through various sensors (\eg accelerometer, swipe patterns), where the initial user effort is often downloading an app, or calibrating input sensors. On the other hand, we define \emph{invisible} schemes to be \dw{} authentication schemes requiring no user involvement at all, neither during set-up nor login. Note that not all \dw{} schemes are invisible, as some require a user to carry out an action; for example, in a \dw{} scheme that fingerprints the access device's accelerometer while at rest~\cite{bojinovmobile2014}, users may need to place their device on a flat surface.

\section{The Mimicry Resistance Dimension}
\label{sec:classi}

User authentication typically relies on \emph{something-you-know} (\ie some secret the user knows), \emph{something-you-have}, or \emph{something-you-are}. Biometric-based \uw{} authentication mechanisms (\ie \emph{something-you-are}) are similar in server-side implementation to \emph{something-you-know}, since biometric data is (preferably) stored as a digital secret. 
Since the transmission path from the user's biometric sensor (\eg fingerprint reader) to the authentication server is often untrusted, or at least less so than the path from the sensor to an authenticating application in \ud{} authentication, 
 an attacker may defeat authentication by simply replaying an exposed secret (\eg a fingerprint). Exposure can occur through, \eg guessing or capture. Most \emph{something-you-know} and \emph{something-you-are} web authentication schemes offer limited resistance to mimicry, since exposure of a user secret typically allows attackers to trivially defeat the scheme.

User authentication via \emph{something-you-have} requires verifying servers to both authenticate the hardware token \emph{and} to verify user possession of the hardware token. Hardware tokens are generally electronic devices (\eg a USB OTP token or smartphone) that can be authenticated by the server using, \eg cryptographic techniques.\footnote{Physically Unclonable Functions (PUFs) do not hide a key, but still have a component that may be exposed albeit harder to reproduce/mimic. See Section~\ref{sec:pufs} for further discussion.} Because authenticating these hardware tokens will almost always rely on secrets, \emph{something-you-have} often boils down to a \emph{something-you-know}, \ie something \emph{the hardware token} knows. Most known variations of hardware authenticator tokens are defeated once that secret is exposed~\cite{bonneau2012quest}, \eg by capture or theft, again providing little to no resistance to mimicry.

\subsection{The Exposure-Mimicry Duality}
\label{sec:duality}
Defeating web authentication typically requires two actions: exposing a token and mimicking certain behavior. That behavior is any action the legitimate user (or device) normally performs while authenticating to a service; for passwords, that behavior is trivially mimicked by simply submitting a static string. A scheme's resistance to compromise thus depends on its ability to independently resist \emph{exposure} and \emph{mimicry}. To evaluate a scheme's resistance across these orthogonal components, we construct a two-dimensional space in Fig.~\ref{fig:mimicrychart} and plot various web authentication schemes on it. A marker (\ie dot) represents a scheme's authentication token. 
Height along the y-axis indicates resilience to exposure; distance from the origin along the x-axis indicates resilience to mimicry. Schemes placed on the chart are explained in Section~\ref{sec:placement}.

\newcommand{\OBCspl}{OBCs (private key leaked)}
\newcommand{\OBCscl}{OBCs (leaking challenge-response)}

\newcommand{\wid}{3.35}
\newcommand{\hei}{3.35}

\begin{figure}
\centering
\begin{tikzpicture}
\begin{axis}[
width=\wid in,
height=\hei in,
ymin=0, ymax=19,
xmin=0, xmax=19,
axis y line=left,
axis x line=bottom,
xlabel={Mimicry\\[0.4em] Resistance},
ylabel=Exposure Resistance,
minor xtick={2,4,8,10,14,16},
minor ytick={1,2,3,4,5,6.4,6.8,7.2,7.6,8,8.4,8.8,9.2,9.6,10,10.4,10.8,11.2,11.6,13,14,15,16,17,18},
xtick={6,12},
ytick={6,12},
xticklabels=,
yticklabels=,
x label style={font=\tiny,at={(1.05,0.125)},text width={1.2cm}},
y label style={font=\tiny,rotate=-90,,at={(0.2,1.02)}},
legend style={draw=none,at={(1.35,0.95)},anchor=north,}
]
\addplot[domain=0.01:18,fill=gray!5,draw=none,samples=250] {sqrt(18^2-x^2)}\closedcycle;
\addplot[domain=0.02:12,fill=gray!15,draw=none,samples=100] {sqrt(12^2-x^2)}\closedcycle;
\addplot[domain=0.03:6,fill=gray!25,draw=none,samples=100] {sqrt(6^2-x^2)}\closedcycle;
\addplot[scatter,mark size=1.5pt,only marks,scatter/@pre marker code/.style={},scatter/@post marker code/.style={},] coordinates{
(0,16)	
(2,11.2)	
(0,10.4)	
(0,9.2)		
(0,4) 
(0,3) 
(0,2) 
(0,1) 
(2,1) 	
(4,4) 	
(0,4) 	
(10,4) 	
(16,1) 	
(16,1)	
(10,1)	
(8,1) 	
(8,1)	
(8,10.4)	
(10,10.4)	
(10,9.2)	
}
node[black,font=\tiny,at={(0,160)},anchor=west]{OTP2, PUF1}
node[black,font=\tiny,at={(20,112)},anchor=west, text width={5cm}]{PassWindow}
node[black,font=\tiny,at={(0,104)},anchor=west]{OTP1}
node[pin=north:,black,font=\tiny,at={(55.5,84)},anchor=west]{OTP3, OTP4}
node[black,font=\tiny,at={(100,104)},anchor=west]{Sound-proof}
node[black,font=\tiny,at={(0,92)},anchor=west]{Social re-auth}
node[black,font=\tiny,at={(0,40)},anchor=west,text width={1.2cm}]{Iris, FP1}
node[black,font=\tiny,at={(0,30)},anchor=west]{Federated}
node[black,font=\tiny,at={(0,20)},anchor=west]{Voice, PCCP}
node[pin=north west:,black,font=\tiny,at={(0,2)},anchor=west, text width={2.8cm}]{Passwords, FP3, L1}
node[black,font=\tiny,at={(100,92)},anchor=west]{PUF2}
node[black,font=\tiny,at={(40,40)},anchor=west, text width={2.8cm}]{FP6}
node[black,font=\tiny,at={(20,10)},anchor=west, text width={2.8cm}]{Cognitive}
node[black,font=\tiny,at={(100,40)},anchor=west, text width={2cm}]{FP2}
node[black,font=\tiny,at={(160,10)},anchor=west, text width={2cm}]{L4, FP4}
node[black,font=\tiny,at={(100,10)},anchor=west, rotate=0, text width={1.3cm}]{L3}
node[pin=205:,black,font=\tiny,at={(75,24)},anchor=west, text width={2cm}]{\ \ \ FP5, L2}
;
\addplot[dashed,opacity=0.4] coordinates{(18,0)(18,18)};
\addplot[dashed,opacity=0.4] coordinates{(12,0)(12,18)};
\addplot[dashed,opacity=0.4] coordinates{(6,0)(6,18)};
\addplot[dashed,opacity=0.4] coordinates{(0,18)(18,18)};
\addplot[dashed,opacity=0.4] coordinates{(0,12)(18,12)};
\addplot[dashed,opacity=0.4] coordinates{(0,6)(18,6)};
\end{axis}

\node at (1.05,-0.65) [align=center] {\tiny{\begin{tabular}{c}Infrequent intercept/\\ capture resistant\end{tabular}} \\[-0.3em] \tiny{} \\[-0.3em] \tiny{H1}};
\node at (3.3,-0.65) [align=center] {\tiny{\begin{tabular}{c}Intercept/capture\\ and reuse resistant\end{tabular}} \\[-0.3em] \tiny{(Spoofable)} \\[-0.3em] \tiny{H2}};
\node at (5.5,-0.65) [align=center] {\tiny{Spoof-resistant} \\[-0.3em] \tiny{\begin{tabular}{c}(Requires\\ Targeted Attacks)\end{tabular}} \\[-0.3em] \tiny{H3}};

\node at (-0.75,1.05) [rotate={90},align=center] {\tiny{V1} \\[-0.3em] \tiny{(Guessable)} \\[-0.3em] \tiny{Negligible-resistance}};
\node at (-0.75,3.3) [rotate={90},align=center] {\tiny{V2} \\[-0.3em] \tiny{(Leakable)} \\[-0.3em] \tiny{Guess-resistant}};
\node at (-0.75,5.5) [rotate={90},align=center] {\tiny{V3} \\[-0.3em] \tiny{\begin{tabular}{c}(Requires\\ Targeted Attacks)\end{tabular}} \\[-0.3em] \tiny{Leak-resistant}};

\end{tikzpicture}
\caption{Exposure resistance and mimicry resistance as two dimensions to measure authentication-scheme resistance to compromise. Along each axis, distance from the origin reflects the scalability of attacks required to defeat a scheme: those plotted closer to the origin (darker regions) can be defeated by attacks that can be scaled with ease (such as online guessing) to break a large number of accounts; defeating schemes plotted further from the origin requires more targeted attacks (such as device theft) that are highly unscalable. Section~\ref{sec:placement} gives detailed explanations of different classes of one time passwords (OTP$n$), device fingerprinting (FP$n$), geolocation (L$n$) techniques, and physically unclonable functions (PUF$n$).}
\label{fig:mimicrychart}
\end{figure}
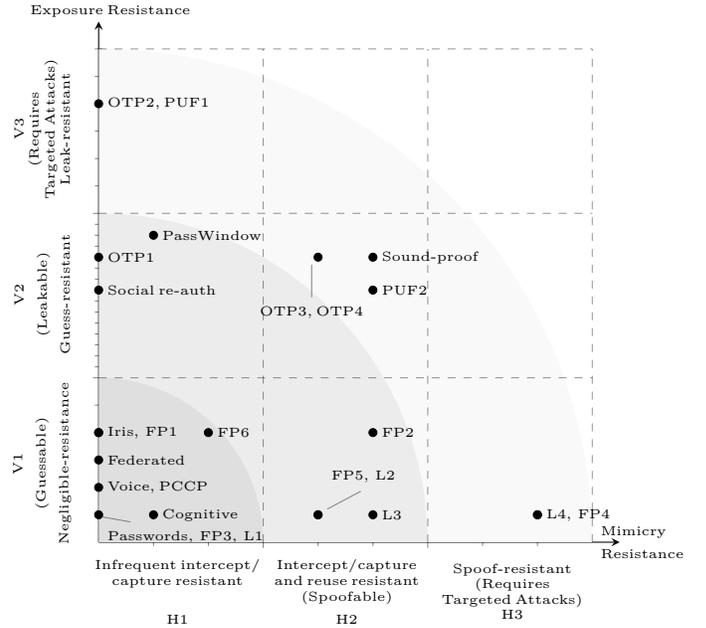

\subsubsection{Vertical axis}
\label{sec:verticalaxis}
The y-axis of Fig.~\ref{fig:mimicrychart} is split into three segments: V1-Negligible-resistance (\ie guessable), V2-Guess-resistant, and V3-Leak-resistant. Guessing a credential is the easiest form of exposure. 
Digital theft (\emph{leak} henceforth) is generally easier than physical theft in that (1) leaks can often be arranged at scale, \eg by phishing, and (2) leaks can be by remote access, \eg an Internet-facing authentication server is subject to attack from anywhere in the world (an adversary need not be in physical proximity).

{\bf V1.} Within the lower-vertical segment, a credential requiring more guesses is placed higher. The guessability of a scheme's credential may depend on several factors. A very weak password, for example, could be easier to guess than a randomly-generated PIN. Unless stated otherwise, schemes are plotted according to their strength in typical scenarios (\eg passwords are assumed to be user-chosen, which limits their resilience to guessing). Guessing attacks are assumed to follow common guessing strategies for \emph{trawling attacks} (capturing as many accounts as possible, often by guessing the most common password across all accounts, then moving down a list of candidate passwords). 

{\bf V2.} The middle-vertical segment comprises schemes with secrets impractical to guess, such as cryptographic keys or randomly-generated passwords with sufficient length to withstand offline guessing attacks. Those may still be leaked (\eg client-side malware). Malware may either (1) steal login credentials for later user impersonation, or (2) remain dormant until a user logs into their account, after which attacker-created transactions can be authorized by the malware on the victim user's device \cite{mannan2011leveraging}. Herein we are interested in evaluating an authentication scheme's resilience to impersonation, not to malicious authorization. Defending against the latter may require mechanisms such as out-of-band authorization for sensitive transactions (which may be independent from the method of user authentication being used). 

We consider within this segment four sublevels, based on the number of sources from which leaks may occur: the human user, the browser, the user device (\eg laptop or smartphone), and a public server whose compromise defeats authentication; \eg a trusted third party of the main authentication server, or a party that also stores the same credential (\eg a same password used across multiple websites). A scheme subject to leaks from all four sources is placed at the lowest of V2's four sublevels; one subject to leaks from any three places it second-lowest, and so on. A scheme not subject to leaks from any of these four sources is placed in V3.

{\bf V3.} Schemes in the upper-vertical segment are those resilient to exposure by digital theft (\eg leak or capture)---thus mostly physical tokens of some sort. Compromising these requires targeted attacks (most commonly physical theft) involving physical proximity to specifically pre-identified users. A scheme's vertical position within V3 varies with vulnerability to theft. A smartphone for example, relatively small in size and carried around more often, is easier to steal than a desktop PC---the latter may require physical break-in to an office and effort to conceal the escape.

{\bf Sorting rationale.} 
The intuition behind arranging the three segments in the above manner follows logically from the trawling attack strategy of maximizing the number of accounts broken into per unit of attack effort. Assuming, for example, that a website does not throttle online password guessing, a good attack strategy is to try guessing passwords. If online guessing fails, an attacker often moves to digital theft (V2 segment on Fig.~\ref{fig:mimicrychart}), \eg stealing credentials via XSS attacks, phishing, or client-side malware. Since random passwords may be leaked but not efficiently guessed, they are harder to expose. Weak passwords or PINs, for example, can be guessed and often also captured,\footnote{An example where a password can be guessed but not leaked would be challenge-response schemes where the password is never typed on the keyboard nor stored anywhere (neither on the server nor any client device), such that the user computes the response from the challenge in their head, or offline using a calculator. That would be resilient to phishing, theft, malware, leaks from verifiers, requires no trusted third parties, and possibly physical observation.} and are rated lower in resistance to exposure. If all forms of leaks (digital theft) fail, the attacker is left with physically stealing, \eg a smartphone or hardware authentication token. These attacks also become gradually harder to scale (for an adversary) in the aforementioned order.

{\bf Relationship to UDS framework.}
As summarized in Table~\ref{tab:udsrelation}, the vertical axis in Fig.~\ref{fig:mimicrychart} addresses the \textsc{S} (Security) benefits in the UDS framework, excluding {\it Unlinkable} and {\it Requiring-Explicit-Consent}.\footnote{These two security benefits from the original UDS framework do not share the focus of Fig.~\ref{fig:mimicrychart} on exposure-resistance. {\it Unlinkable} is a privacy benefit and {\it Requiring-Explicit-Consent} relates to malicious authorization, \eg a malicious RFID-based card reader embedded in a sofa that authorizes a transaction without user knowledge~\cite{bonneau2012quest}.} The UDS framework did not intend to provide an overall summary rating for a scheme, as that would require subjective weighting of the usability, security, and deployability benefits (which may be context-dependent). In contrast, we use our sorting rationale (as discussed above) in conjunction with the UDS security benefits to rate a scheme's overall resistance to exposure; for example, if a scheme only fails to provide \emph{Resilient-to-Unthrottled-Guessing} but provides the remaining eight security properties, it is placed near the bottom of the vertical axis within V1 (despite having a virtually full row of security bullets). Passwords fail to provide \emph{Resilient-to-Throttled-Guessing} and \emph{Resilient-to-Unthrottled-Guessing} and are therefore placed in V1, despite providing \emph{Resilient-to-Physical-Theft} (which corresponds to our highest exposure resistance category, segment V3).

\begin{table}
\caption{Original UDS security benefits \cite{bonneau2012quest} evaluating schemes by susceptibility to exposure, and relation to new framework. This table lists the properties a scheme must offer to move to the next-higher vertical segment in Fig.~\ref{fig:mimicrychart}; \ie both S3 and S4 must be offered to be placed in V2; all of S1, S3-S7 and S9 must be offered to be placed in V3. The benefits are listed in descending order of the scalability of carrying out the corresponding attack; \eg guessing attacks (S3) are highly scalable, but physical theft (S8) is highly unscalable.}
\label{tab:udsrelation}
{
\centering
\begin{tabular}{l|l@{}l}
\toprule

\multicolumn{1}{c|}{\textsc{UDS Security Property}}
&
\multicolumn{2}{c}{\textsc{Segment}}\\\hline

S3. {\it Resilient-to-Throttled-Guessing}	& \multirow{2}{*}{Lower} & \multirow{2}{*}{\ (V1)}\\
S4. {\it Resilient-to-Unthrottled-Guessing}\\\hline

S1. {\it Resilient-to-Physical-Observation} & \multirow{5}{*}{Middle} & \multirow{5}{*}{\ (V2)}\\
S5. {\it Resilient-to-Internal-Observation}\\
S6. {\it Resilient-to-Leaks-from-Other-Verifiers}\\
S7. {\it Resilient-to-Phishing}\\
S9. {\it No-Trusted-Third-Party$^{*}$}	\\\hline

S2. {\it Resilient-to-Targeted-Impersonation$^{\dagger}$}& \multirow{2}{*}{Upper} & \multirow{2}{*}{\ (V3)}\\	
S8. {\it Resilient-to-Physical-Theft}	\\	

\bottomrule 
\end{tabular}\\[0.7em]
}
$^{*}${\footnotesize Additional third parties increase chances of leaks from public servers. From the adversary's perspective, this property is similar to {\it Leaks-from-Other-Verifiers}. See inline for details.}\\
$^{\dagger}${\footnotesize S2 encompasses a variety of targeted attacks, \eg deceiving a human acting as a trusted party as in social re-authentication~\cite{brainard2006fourth}, locating a user's password written on a post-it note, or lifting fingerprints from a doorknob~\cite{bonneau2012quest}. The success of these attacks require attackers to be in physical or logical proximity (\eg social engineering attacks against social re-authentication) to victims. The difficulty in scaling such targeted attacks is similar to attacks involving physical theft (S8). These differ from more scalable attacks such as educated guessing based on information scraped \emph{en-masse} from social media---these are covered by the more general guessing attacks (S3 and S4).}
\end{table}

\textbf{Method for Plotting Schemes (Vertical Axis).} Schemes are plotted along the vertical axis (see Section~\ref{sec:placement} for placement rationale) based on their security benefits offered. For schemes already evaluated in the UDS paper~\cite{bonneau2012quest} we use the rating given therein. The following criteria (based on the mappings in Table~\ref{tab:udsrelation}) are used to decide which of the three vertical segments (V1, V2, V3) the scheme will fall into:
\begin{enumerate}
	\item Schemes subject to guessing attacks (\ie not offering both S3 and S4) are placed in V1.\footnote{When evaluating specific implementations of authentication schemes, resistance to guessing attacks can be more precisely calculated (\eg based on the specific throttling techniques used); V1 could then be divided into more granular sublevels based on the number of guesses required to break a scheme. Schemes breakable with zero guesses would be plotted at the origin, and schemes breakable with an offline guessing attack of just under $10^{20}$ guesses (the approximate threshold at which a scheme would be considered to be resistant to offline guessing attacks \cite{florencio2014administrator}) would be plotted at the upper boundary of V1.}
\item Schemes offering S3 and S4, but lacking any of S1, S5-7, or S9, are placed in V2.
\item Schemes offering all of the above benefits are placed in V3.
\end{enumerate}
Moreover, each vertical segment has sublevels that are determined based on the number of security benefits captured within each segment, as specified in Table~\ref{tab:udsrelation}. Each segment has precisely $\sum^{N+1}_{i=2}{i}$ sublevels, where $N$ is the number of benefits associated with that segment; thus, V1 and V3 each have five sublevels, and V2 has 14 sublevels. Each scheme is assigned to a sublevel such that a security benefit offered (\ie represented by a filled-circle ``$\bullet$'' in Table~\ref{tab:prims}) has a greater score than any number of partially-offered benefits (\ie represented by an empty-circle ``$\circ$'' in Table~\ref{tab:prims}). For example, a scheme that offers all V1 benefits but no V2 benefits would be placed at the first sublevel of V2; a scheme that partially offers four V2 benefits would be placed at the fifth sublevel of V2; a scheme that offers one V2 benefit and partially offers one V2 benefit would be placed at the seventh sublevel of V2; and so on.\footnote{To avoid visual ambiguity, no sublevels are plotted at the boundary between any two segments, namely the V1-V2 and V2-V3 boundaries on the vertical axis or the H1-H2 and H2-H3 boundaries on the horizontal axis. Therefore, no schemes are placed at these boundaries.}

\subsubsection{Horizontal axis}
\label{sec:horizontal}
The x-axis (Mimicry Resistance) reflects burdens on the attacker to mimic the verifier-expected behavior, \emph{after} exposing a scheme's credential. Unfortunately, the security of most authentication schemes proposed to date rely on some form of resisting exposure, with compromise complete once an underlying secret token is revealed. This includes passwords, \uw{} physical biometrics (see Section~\ref{sec:conveyor}), and poorly managed/generated private keys~\cite{heninger2012mining}. 
If leaking a credential does not allow an attacker access by direct replay of the credential, the scheme exhibits some degree of mimicry resistance, and gains horizontal distance from the origin in Fig.~\ref{fig:mimicrychart}.

We split the horizontal axis into three segments: H1-Negligible-resistance (\ie easily replayable), H2-Replay-resistant, and H3-Spoof-resistant. In Section~\ref{sec:added}, we expand the UDS framework by six new benefits; three of these benefits (M2, M3, and M4) correspond to the aforementioned horizontal segments (H1, H2, and H3) as specified by Table~\ref{tab:mimicrelation}.

\begin{table}
\caption{Three new UDS-type security properties (given in Section~\ref{sec:added}) directly related to mimicry. A scheme must achieve these to move further right across the horizontal segments in Fig.~\ref{fig:mimicrychart}. Example: Both M2 and M3 must be achieved to be placed in H3.}
\label{tab:mimicrelation}
\centering
\begin{tabular}{@{}l|l@{}l@{}}
\toprule

\multicolumn{1}{c|}{\textsc{New UDS Security Properties}}
&
\multicolumn{2}{c}{\textsc{Segment}}\\\hline

M2. {\it Resilient-to-Infrequent-Capture-or-Intercept}	& \multirow{1}{*}{Left-most} & \multirow{1}{*}{\ (H1)}\\

M3. {\it Verifies-Non-Static-Information} & \multirow{1}{*}{Middle} & \multirow{1}{*}{\ (H2)}\\

M4. {\it Resilient-to-Spoofing}& \multirow{1}{*}{Right-most} & \multirow{1}{*}{\ (H3)}\\

\bottomrule 
\end{tabular}
\end{table}

{\bf H1.} If an attacker can simply submit or replay a credential after exposing it, the scheme provides no resistance to replay and thus gains no horizontal distance along the Mimicry Resistance dimension (\ie lies on the vertical axis). For example, replaying a captured password is trivial; thus, passwords lie directly on the vertical axis in Fig.~\ref{fig:mimicrychart}.

In some cases, there is no clear-cut answer as to whether exposing information directly helps the attacker gain access. For example, answers to personal knowledge questions based on recent account activity (\eg \textsl{who was the last person you emailed?}) likely remain unchanged for a few hours. Capturing an answer allows the attacker to login only within that window. Such schemes are given some horizontal distance from the origin, but remain within H1. Another example is password expiration policies~\cite{zhang2010security,chiasson2015quantifying} forcing resets every, \eg one hour, which would provide some mimicry resistance (albeit highly unusable), and placed within H1 on Fig.~\ref{fig:mimicrychart}.

Some challenge-response schemes are also candidates for H1; whether or not a captured response directly allows the attacker to compromise the account is conditional on whether the captured response remains valid. Schemes where the set of challenge-response pairs is finite and relatively small, or when capturing a handful of challenge-response pairs suffices for defeating authentication, would thus fall within H1. Cognitive schemes like Weinshall~\cite{weinshall2006cognitive}, and some challenge-response schemes such as PassWindow \cite{passwindow}, are prominent examples.
Other challenge-response schemes are placed in H2.

{\bf H2.} After a credential is leaked, schemes that either require additional attacker actions beyond simply replaying a string or conducting a straightforward operation (\eg cryptographic signing), or where attackers have a limited time window (\eg $\textless$ 2 mins) in which the credential can be used, are considered replay-resistant and placed within H2. An example is DNS resolver fingerprinting (FP5 in Fig.~\ref{fig:mimicrychart}), which determines the DNS resolver used by the client (see Section~\ref{sec:devicefp}); knowing the resolver's address is insufficient to defeat authentication, as the attacker must also be able to use it to resolve domain names.

One-time password (OTP) over SMS is another example scheme placed within H2, as it enables the attacker to compromise the account only if the user has not already used the OTP. An attacker that captures the OTP in clear text~\cite{signalinglink} must use it before the legitimate (victim) user. 
Additionally, the server may set a 2-minute time window where an OTP token is valid for usage, and expires afterwards. 
Such schemes would be placed within H2 on Fig.~\ref{fig:mimicrychart}.

{\bf H3.} We place within H3 any scheme that requires additional equipment and/or systems (\eg hardware chip manufacturers, or large scale distributed botnets) to mimic the behavior that the server measures and expects from the legitimate user. An example is robust location verification (see L4 in Section~\ref{sec:geogeo}).

Because spoof-resistant schemes rely on measurements of various phenomena (\eg a user behavior or habit), the measurement process must typically allow some degree of tolerance to account for (1) imperfections in the measuring apparatus, or (2) the phenomena's natural instability over time. 

For (1), consider an example location-based authentication scheme \cite{van_oorschot_identity_2005}, where a server grants access to a user only if the user is at an expected geographic location. Ideally, the user's location would be identified to such a high granularity that no two human beings could exist at the same \{latitude, longitude, altitude\} coordinates. 

For (2), again using location-based authentication as an example, a user's daily commute \cite{jakobsson_implicit_2009} from one geographic location to another (\eg work to home) makes it impractical to grant account access only when the user is geographically present in, \eg a 1 $m^2$ area, even if the geolocation mechanism being used allows for such high accuracy. Geolocating users with courser granularity (\eg city-level) would thus seem more practical (\eg to avoid false rejects) for generic location-based authentication purposes. 

The degree of ``fuzziness'' (\ie lack of precision) introduced while measuring a user's behavior in spoof-resistant schemes is thus likely to result in \emph{false accepts}, \ie falsely accepting another user as the intended one. In practice, some websites use IP-address based location look-ups~\cite{poese2011ip} to implement location-aware authentication, which returns locations at the city or state level. Thus, all attackers physically present in the user's city (or even a legitimate user mistyping the username with no malicious intent, \ie if geolocation is used as a stand-alone authentication scheme) are falsely accepted.

Note the distinction between false accepts (a result of imprecision) and mimicry resistance (a result of resistance to attack): A scheme that uses high-precision GPS coordinates (\eg within a radius of $1m$) reported by the user browser (\eg L1, Section~\ref{sec:geogeo}) may offer low false accept rate, but offers no mimicry-resistance since attackers can use browser extensions that report forged coordinates.

False accepts are not directly reflected in the Mimicry-Resistant dimension (though we capture them in the evaluation framework in Section~\ref{sec:added}). They are indirectly captured however by a scheme's position vertically. 
This is because resilience to guessing attacks and resilience to false accepts are both related to the size and distribution of the credential space (\eg cryptographic keys with sufficient length and chosen uniformly at random are not subject to guessing attacks and false accepts).

\textbf{Method for Plotting Schemes (Horizontal Axis).} Schemes are plotted along the horizontal axis based on their mimicry-resistance benefits offered. The following criteria (based on the mappings in Table~\ref{tab:mimicrelation}) are used to decide which of the three horizontal segments (H1, H2, H3) the scheme will fall into:
\begin{enumerate}
	\item Schemes that do not offer benefits M2, M3, and M4 offer no mimicry-resistance, and thus are placed on the vertical axis (\ie with zero horizontal displacement).
	\item Schemes offering M2 are placed in segment H1.
	\item Schemes offering M3 are placed in segment H2.
	\item Schemes offering M4 are placed in segment H3.
\end{enumerate}

Since H1, H2, and H3 are each associated with a single benefit (see Table~\ref{tab:mimicrelation}), each horizontal segment has two sublevels, where the first sublevel represents a partially-offered benefit and the second sublevel represents a fully-offered benefit. 

\subsection{Interpreting Attack Scalability}
\label{sec:scalability}

Figure~\ref{fig:mimicrychart} illustrates the relative resilience of schemes against attackers that aim to maximize the number of accounts broken into (\ie trawling attacks). Along each individual axis, schemes further from the origin are less susceptible to scalable attacks. Scalable attacks are defined as attacks that can be scaled to break into a large number of accounts. Non-scalable attacks are typically highly targeted (\eg physical device theft); such attacks impose a higher cost on the attacker in terms of time, effort, and/or money, as a function of the total number of accounts targeted.

As our evaluation is qualitative, it is not intended that absolute distance from the origin be a quantitative measure suitable for comparing schemes. However, the resilience of two schemes (\ie scalability of the attacks that a scheme can withstand) can be compared if (1) they share one of the two coordinates (\ie identical $x$- or $y$-coordinates), or (2) one of the schemes is further from the origin along both its $x$- and $y$-coordinates. When comparing any two schemes that satisfy one of these conditions, a scheme that offers both high exposure-resistance (\ie high $y$-value in Fig.~\ref{fig:mimicrychart}) and high mimicry-resistance (\ie high $x$-value in Fig.~\ref{fig:mimicrychart}) is more costly (less scalable) to attack, since it requires the attacker to successfully mount both an exposure attack and mimicry attack, as elaborated on below.

Along the vertical (exposure-resistance) axis, schemes most vulnerable to scalable attacks are those where the attacker can guess user credentials (V1). Attacks against schemes requiring information theft (\eg via software vulnerabilities or malware) are less scalable (V2), and schemes requiring physical device theft (V3) are the least scalable (\ie most costly) to attack.

Along the horizontal (mimicry-resistance) axis, attacks against schemes where an exposed secret can be used as-is by the attacker (see Section~\ref{sec:duality}) are most scalable; attack scalability decreases for schemes that increase the attacker burden (in terms of hardware costs or control of network infrastructure) for mimicking account-holders beyond a simple replay attack or capture of a static secret (see Table~\ref{tab:mimicrelation}).

\subsection{Scheme Placement on Figure~\ref{fig:mimicrychart}}
\label{sec:placement}

We use the mapping method of Section~\ref{sec:verticalaxis} (``Relationship to UDS framework'') to position a representative subset of schemes along the exposure-resistance axis using their previously assessed (\cf~\cite[p.11]{bonneau2012quest}) security benefits. Upon evaluating the mimicry-resistance of these schemes, we find they offer little to no resistance to mimicry as we explain below.

PassWindow~\cite{passwindow} is a visual crypto technique whereby the user overlays a transparency card on the screen to visually read and enter a 4-digit pass-code, which proves possession of the user's card. It was shown~\cite{nettle2009passwindow} that observing 20--30 challenge/response pairs can leak the card's secret; we thus place the scheme within H1. Likewise, cognitive schemes, including GrIDsure~\cite{jhawar2011make}, Weinshall~\cite{weinshall2006cognitive}, Hopper Blum~\cite{hopper2001secure}, and Word Association~\cite{smith1987authenticating} are placed in H1 since the schemes can be compromised after few observations (\eg Weinshall can be compromised after about seven observations \cite{golle_cognitive_2007}).

In social re-authentication~\cite{brainard2006fourth}, a trusted friend can vouch for a user who, \eg has lost their credentials. An attacker may capture a vouch code (\eg via malware on a trusted friend's device) and simply replay it to authenticate as the victim; we thus place it in H1.

Federated schemes, including OpenID~\cite{recordon2006openid}, Microsoft Passport~\cite{kormann2000risks}, Facebook Connect, and BrowserID~\cite{hanson2011federated}, offer no mimicry-resistance if password authentication is used, and we thus place them in H1. Their position on the chart would change if the identity provider utilizes mimicry-resistant schemes. Persuasive Cued Click-Points (PCCP)~\cite{chiasson2012persuasive} and biometric schemes on Fig.~\ref{fig:mimicrychart} (Iris and Voice) are also placed in H1, since captured biometric samples are typically replayable~\cite{bonneau2012quest}.

We explore and discuss example schemes which do provide certain mimicry-resistance, and position them along the two axes (the remaining schemes on Fig.~\ref{fig:mimicrychart}). To avoid redundancy, we only explain the horizontal position of these schemes in this section; see Section~\ref{sec:added} for their vertical position.

\subsubsection{Geolocation (selected methods)}
\label{sec:geogeo}

We review four broad classes of Internet geolocation techniques that can be used for location-based authentication, placing a marker for each class on Fig.~\ref{fig:mimicrychart} to rate their ability to resist compromise. Location-based authentication typically requires server-side storage of the user's expected location (\eg city-level, nation/state level, or latitude/longitude coordinates, depending on the geolocation method employed) in plain text, hashed, or cloaked~\cite{damiani2011fine} to a certain degree to preserve users' privacy. Any of the four classes of geolocation methods described below may be used by a verifying server to obtain and/or verify the user's current location at the time of authentication, and grant access if the user is verified to be at the location expected by the server.

\textbf{L1: GPS and WPS.} GPS and WiFi Positioning System (WPS) geolocation are commonly used in practice. WPS uses multi-lateration based on the signal strengths between the device and nearby WiFi access points with known locations. These techniques are usually selected by the user's browser in the W3C geolocation API~\cite{popescu2010geolocation}, whereby the browser obtains the coordinates from the device's GPS driver, or from a location provider after submitting a list of nearby WiFi access points and their signal strengths to the location provider. L1 gains no depth across the Mimicry Resistance dimension (see Fig.~\ref{fig:mimicrychart}), since techniques rely on browser-reported information that can be substituted and replayed by an attacker, so knowledge of the user's location is sufficient to break authentication.

\textbf{L2: IP-address based tabulation.} Tabulation-based geolocation service providers such as \texttt{Maxmind}\footnote{\url{https://www.maxmind.com/}} and \texttt{ipinfo}\footnote{\url{http://ipinfo.io/}} maintain lookup tables, which map IP address blocks to cities and countries, possibly through publicly available information such as IP address registries (\eg \emph{whois}\footnote{\url{https://www.whois.net/}}) and the geography of IP address allocation. Geolocation based on IP address table lookups is unreliable due to outdated entries~\cite{poese2011ip}, and is evadable through use of middleboxes and virtual private networks (VPNs)~\cite{muir2009internet}. L2 is rated slightly more resilient to compromise than forgeable GPS/WPS coordinates, and is placed within H2; it is not in H3 since even an attacker unable to forge source IP addresses may use public HTTP proxies or bots in close proximity to the user. 

\textbf{L3: Measurement-based geolocation.} In this class, network measurements, such as Round-Trip Times (RTTs), are conducted from a set of landmarks (\eg cloud-based or CDN servers) to the target user, and are then mapped to geographic distances using a pre-calibrated delay-to-distance function. The user's location is estimated through multi-lateration, relative to the landmarks' locations. Examples include Spotter~\cite{laki2011spotter} and CBG~\cite{gueye2006constraint}. Other proposals have suggested mapping the network topology for higher accuracy~\cite{li2013ip}. To date, measurement-based geolocation can achieve an accuracy on the order of a few tens of meters. Manipulating L3 requires more advanced techniques than simply submitting forged coordinates, but can be achieved with enough knowledge of the network topology and the landmarks/verifiers being used~\cite{abdou2017accurate,gill2010dude}; they are limited to H2.

\textbf{L4: Robust location verification.} Some techniques are designed to verify location information obtained by other Internet geolocation techniques (\eg L1-L3, above), and/or are designed to be resilient to common adversarial manipulation tactics. The result of a preliminary geolocation is treated as an asserted location (analogous to a username asserting identity), to be verified by a measurement-based proof (analogous to proof of knowledge of a secret). Examples of such schemes include Client Presence Verification (CPV)~\cite{cpvtdsc} (see Section~\ref{sec:added}) and Trusted Platform Module (TPM)-supported GPS drivers~\cite{TGVisor}; the former cryptographically protects network delay measurements used for verifying location assertions, and the latter communicates coordinates securely. We call such techniques robust location verification and rate them as spoof-resistant (H3) because manipulation requires attackers to expend more effort than simply reporting a false location; successfully spoofing legitimate client locations requires using specialized proxy machines (for attacks that ``co-locate'' using a machine nearby to the victim) or GPS satellite signal-spoofing devices~\cite{tippenhauer2011requirements}.

\subsubsection{Device Fingerprinting} 
\label{sec:devicefp}

Device fingerprinting refers to techniques by which a server collects information on a device's hardware/software configuration for the purpose of identification~\cite{eckersley2010unique}. From 29 methods recently surveyed~\cite{fp2016acsac}, we derive six representative categories for evaluation.

\textbf{FP1: System parameters/preferences.} This class includes software and hardware information about the device to be authenticated, provided by the web browser's JavaScript API (\eg the \textsf{navigator} and \textsf{windows} JavaScript BOM objects), such as operating system version, screen resolution, time zone, system language, and supported WebGL capabilities. FP1 lies on the y-axis, since it can simply be mimicked by replaying the information.

\textbf{FP2: Audio and canvas challenge/response.} This class includes two techniques that fingerprint the client's graphics and audio subsystems, respectively. HTML5 canvas fingerprinting renders a variety of text and graphics in an HTML5 canvas on the client's browser, which results in subtly different images (\eg due to differences in anti-aliasing or font smoothing) depending on the graphics driver and hardware on the device being fingerprinted. Audio processing fingerprinting leverages the HTML \textsf{AudioContext} API that provides real-time frequency- and time-domain analysis of audio playback, mainly used for creating audio visualizations. Playing the same sound on different devices results in subtly different waveforms, depending on the sound driver and hardware. To improve resistance to replay attacks, these two techniques can be used in a challenge-response scheme, where the server can store the client's responses to many different challenges~\cite{bursztein2016picasso}, and are thus placed within H2. 

\textbf{FP3: Hardware sensors.} This class, suitable to fingerprint smartphones, leverages the inherent variation in the manufacturing and factory calibration of typical smartphone sensors, such as accelerometer and speaker-microphone systems. Similar to FP1, FP3 lies on the y-axis since information can simply be replayed.

\textbf{FP4: Clock skew.} The server uses TCP timestamps to measure the clock skew of the device being fingerprinted, which differs across devices due to manufacturing variation. FP4 is placed within H3, since clock skew spoofing attacks are highly sophisticated; \eg Arackaparambil \etal show how timestamp manipulation can be detected to identify rogue wireless LAN access points \cite{arackaparambil2010}. However, clock-skew spoofing over a WAN connection is not well-studied, and therefore this rating may change subject to further study and experimentation.

\textbf{FP5: DNS resolver.} The server determines a client's DNS resolver(s) by presenting to the client a page that contains a reference to a randomly-generated (non-existent) subdomain, triggering a client DNS lookup; as a result the server will receive a DNS query from the client's resolver. The DNS resolver's IP address serves as a fingerprint. FP5 provides partial replay-resistance but can be defeated via the use of proxies, similar to L2. In some cases, organizations run their own DNS servers, which resolve domain names only to machines within the department's network. To use a victim's DNS resolver, an attacker would need to be able to resolve domain names using the organization's private DNS server. Since identifying the server is not enough (and is easier than using the organization's DNS), FP5 is placed within H2.

\textbf{FP6: Protocol-based fingerprinting.} This class includes schemes that glean information from \mbox{network-,} transport-, and application-layer protocol fields. For example, the TLS library of the client device can be fingerprinted using the \textsf{Client Hello} packet received during the handshake sequence, which includes information such as the device's supported TLS version, supported ciphersuites (and their order of presentation), compression options, and list of supported extensions (along with associated parameters such as elliptic curve parameters). FP6 lies within H1, since it is susceptible to mimicry, but attacks are less scalable than simply replaying a static string; OS- or library-level modifications may be required.

 \subsubsection{OTP Schemes}
\label{sec:otps}

One-time password (OTP) schemes generate short-lived credentials, often used as a second factor alongside conventional passwords. Depending on implementation (four follow), an attacker may aim to capture either the seed or a challenge-response pair.

\textbf{OTP1: OTP mobile apps.} This class (\eg Google Authenticator) generates OTPs to be manually typed into a user's access device, using a combination of a locally-stored shared secret and either the current time (TOTP \cite{totp}) or a counter (HOTP \cite{hotp}). With malware on the device, the attacker can capture the locally-stored shared secret; because this directly enables the attacker to compromise the account, OTP1 provides no resistance to mimicry, and is placed on the vertical axis of Fig.~\ref{fig:mimicrychart}. Note: A variant of this class is a mobile app that stores a public-private key pair, \eg as used by Duo Security~\cite{duosecurityapp} and Google Prompt (a more recent iteration of Google two-step verification~\cite{googleprompt}). When the user types in their password, the server sends a cryptographic challenge to the mobile app. The mobile app then requests user consent, via simply tapping a button, to send the cryptographic response to the server to complete the user authentication process. The advantage of this approach compared to OTP1 is that it is more efficient to use. The disadvantages are that the mobile app requires an Internet connection, and there is a chance of the user accidentally consenting to a malicious authentication attempt.

\textbf{OTP2: OTP USB tokens.} This class (\eg FIDO U2F keys~\cite{dongleauth}) is similar to OTP1, but requires less effort since the user can press a button on the hardware token to automatically enter the OTP into a browser window. Assuming hardware tokens can resist malware, it becomes relatively challenging for an attacker to capture challenge-response pairs. Similar to OTP1, the attacker can thus target the seed, \ie hardware token theft. This gives the attacker direct account compromise, and therefore is placed on the vertical axis (\ie no horizontal distance from the origin) in Fig.~\ref{fig:mimicrychart}.

\textbf{OTP3: SMS OTP.} The server sends a randomly-generated OTP (\ie no reliance on a shared secret/seed) to the user via SMS. Contrary to the previous two OTP classes, the seed here is only stored on the server, which makes it harder to capture. The attacker can however capture an OTP in transit, \eg by exploring weaknesses in the cellular network~\cite{signalinglink}, but will be required to use it before it expires (and before the user uses it). Since the time window for a successful attack is limited, OTP3 is placed further (horizontally) from the origin, in H2 on Fig.~\ref{fig:mimicrychart}.

\textbf{OTP4: E-mail OTP. } The server sends a randomly-generated OTP (again no reliance on a shared secret/seed) to the user via e-mail. Because an attacker can capture an OTP, \eg via malware on the user's machine, its mimicry-resistance is similar to OTP3.
 \subsubsection{PUFs}
\label{sec:pufs}
Physically Unclonable Functions (PUFs) are hardware modules (typically manufactured into silicon-based chips) that leverage an underlying unique physical structure to generate challenge-response pairs; ideal PUFs are impossible to clone, since the unique structure of each individual PUF results from manufacturing variation \cite{yu2016pervasive}. For our evaluation herein, we assume the use a challenge-response protocol (as described by Yu \etal~\cite{yu2016pervasive}) in conjunction with PUFs that are built into the user's device (as opposed to, \eg a USB-based token).

\textbf{PUF1: Strong PUFs.} These theoretically generate an endless supply of challenge-response pairs, allowing verifying servers to store any number of challenge-response pairs (\eg at user account creation) to be used for user authentication. Several implementations of strong PUFs are available \cite{ruhrmair2014pufs}; \eg some PUFs generate challenge-response pairs by shining a laser on a scattering object at selected angles and points of incidence \cite{pappu2002physical}. A major challenge in developing strong PUFs is to avoid susceptibility to model-building attacks that collect challenge-response pairs and apply machine-learning algorithms to build a mathematical model of the PUF \cite{delvaux2014}. PUF1 lies in V3, since the only security property it lacks from Table~\ref{tab:udsrelation} is \emph{Resilient-to-Physical-Theft}; device theft allows an attacker to directly impersonate the user (\ie defeat authentication), and therefore no mimicry-resistance is offered.

\textbf{PUF2: Weak PUFs.} These can only generate a limited number of challenge-response pairs. They are suitable for authentication when augmented with appropriate mechanisms, \eg when restricted to only responding to challenges sent from a single trusted verifier over an authenticated channel \cite{yu2016pervasive}; this limits an attacker's ability to exhaustively capture all possible challenge/response pairs to mount a mimicry attack. PUF devices are commercially available, \eg for device authentication and cryptographic key storage \cite{intrinsicid}. PUF2 is placed within H2---it lacks \emph{Resilient-to-Internal-Observation} from Table~\ref{tab:udsrelation}, but a successful mimicry attack can be mounted by building a mathematical model of the PUF only after capturing a sufficient number of challenge/response pairs (\eg via man-in-the-middle); however, since every PUF is unique, the mathematical model would be device-specific. Counterintuitively, PUF2 exhibits higher mimicry-resistance than PUF1---however, recall that the strength of a scheme is determined by both exposure- and mimicry-resistance. While PUF2 can be mimicked by a determined and targeted attacker even without device posession, a successful attack against PUF1 requires device posession (thereby obviating the need for mimicry), justifying PUF1's higher placement within V3.
 
\subsubsection{Sound-Proof}
\label{sec:soundproof}

Sound-Proof~\cite{karapanos2015sound} is an authentication scheme that determines whether the user's smartphone and access device (\ie another device from which the user is authenticating to access their online account) are in close proximity. The access device records ambient sound from the microphone and transmits it to the smartphone (via the Internet), which compares with the sound recorded by its own microphone. A match indicates that both devices are in an identical noise environment, and therefore likely in close proximity, resulting in the smartphone sending a cryptographically-signed assertion to the web server to approve the user authentication.

We include Sound-Proof in our evaluation as it provides a degree of mimicry-resistance if a trusted channel is established between the smartphone and web server, \eg by means of a TPM (via a trusted execution environment and secure key storage). A TPM-augmented Sound-Proof would address attacks where, \eg the attacker can steal the key via root-privileged malware and use it to sign assertions.  
Note that while Sound-Proof was originally proposed and evaluated as a second factor alongside passwords, we place it on Fig.~\ref{fig:mimicrychart} as a single-factor scheme (without passwords)---Section~\ref{sec:insights} explains our rationale.

Although a TPM-based implementation may render it impractical for an attacker to steal the Sound-Proof app's cryptographic key from the user's smartphone and thus gain permanent access to the account, the attacker may still use a malicious app on the user's device to relay recorded audio. For example, the attacker can reproduce the ambient sound from the environment in which the user's smartphone is currently located, \eg by leveraging malware on the user's smartphone or a nearby device to record and relay the ambient sound to the attacker. 
Sound-Proof is thus placed in H2 because this process is more sophisticated than simply replaying a static token. It is not placed in H1, since repeated (physical or internal) observations do not seem to help the attacker gain permanent access---the authors show~\cite{karapanos2015sound} that the scheme is resilient to guessing attacks even when the attacker knows the user's environment (\eg by using typical background noise from a Starbucks coffee shop), and that attacks are also made more difficult by requiring the sound files recorded by both devices to be timestamped (with NTP-synchronized clocks).

\subsection{Further Insights}
\label{sec:insights}

We briefly discuss further insights from our analysis of mimicry resistance, and the relative strengths of schemes plotted in Fig.~\ref{fig:mimicrychart}.

{\bf Multi-factor authentication.} Fig.~\ref{fig:mimicrychart} should be used to evaluate individual authentication schemes. To evaluate a multi-factor scheme, the scheme must be broken down to its constituent factors, each represented individually as an independent marker. 
Two independent markers can then be combined as a two-factor scheme as follows: the (\emph{x,y}) position of the resultant (combined) scheme is the greater of both x-values and both y-values. That position should however be carefully interpreted. 
For example, although combining L4 with OTP2 would result in a marker in \{H3,V3\} (see Fig.~\ref{fig:mimicrychart}), an attacker capable of physically stealing the device to defeat OTP2 will already be geographically co-located with the user, and thus need not expend any additional effort to defeat L4.

{\bf Lack of schemes in top-right corner of Fig.~\ref{fig:mimicrychart}.} None of the schemes analyzed have strong resistance to both mimicry and exposure---see the three empty squares in the top-right (\{H2,V3\}, \{H3,V2\}, \{H3,V3\}). Schemes in this region would strongly resist scalable attacks. This motivates combining complementary schemes, as discussed immediately above.

{\bf Bands versus markers.} Instead of marker representation, schemes on Fig.~\ref{fig:mimicrychart} could be represented using bands (\ie lines/curves or shaded areas). For example, passwords could be represented by a vertical band from V1 (user-chosen passwords) to somewhere in V2 (system-assigned passwords that are resilient to guessing attacks, but still subject to leaks). Such representation can help identify how different implementations of a scheme may alter security. We chose the simpler representation as it is easier to interpret, and to avoid cluttering the chart with intersecting bands.
 
\section{Comparative Evaluation}
\label{sec:added}
In Table~\ref{tab:prims}, we evaluate selected classes of invisible and mimicry-resistant authentication schemes (selected baseline schemes are also included for comparison). We augment the original 25 usability, deployability, and security properties (benefits) of the UDS framework~\cite{bonneau2012quest} with six new properties relevant to schemes that are invisible and have mimicry-resistance. These new properties were not present in the original UDS paper. The original UDS properties (see Appendix~\ref{sec:uds}) and our new properties defined below are italicized when referenced herein. The new properties are now described:

\textbf{U9. No-False-Rejects } is a usability benefit concerning authentication failures resulting from \emph{system} error (\eg due to measurement error). False rejects arise due to \emph{fuzzy} or non-binary matching functions employed by some mimicry-resistant and/or invisible authentication schemes, and are thus relevant to penalize (by withholding this benefit) schemes that suffer from false rejects (\eg measurement-based Internet geolocation methods). This differs from \emph{Infrequent-Errors} (U7, Appendix~\ref{sec:uds}), wherein authentication may fail due to \emph{user} action (\eg incorrectly typing a password) or attempts to authenticate under unusual circumstances (\eg from unexpected locations).

\textbf{U10. Easy-to-Change-Credentials} is a usability benefit for schemes where a user may easily change credentials (\eg in event of a server database leak). Intuitively, since invisible schemes require no user action upon login or account set-up (see Section~\ref{sec:conveyor}), these schemes rely on remotely (\cf \dw{}) observing habitual user/device attributes and/or behaviours, and transparently verifying these upon login. Changing these credentials (regardless of the reason for changing) is likely to impose user burden, \ie changing physical habits and/or habitual behaviors. This new property (U10) allows appropriate penalization for such schemes. Note that in contrast, \emph{Easy-Recovery-From-Loss} (U8) reflects how easily a user can recover from a credential loss (\eg forgotten password). Credential loss requires a fallback mechanism to verify the user's identity; changing credentials does not, since the user remains in possession of valid credentials. \emph{Easy-to-Change-Credentials} is inherent to the authentication mechanism itself, whereas \emph{Easy-Recovery-From-Loss} may also depend on the fallback mechanism.

\textbf{M1. No-False-Accepts } is a mimicry-related (hence, the `M' index) security benefit of schemes that have a sufficiently large credential space and/or measurement precision such that two different sets of credentials (\eg iris scans from two different individuals) are always distinguishable. False accepts may include both non-malicious users and attackers mistaken for legitimate users, \eg due to close proximity in a geolocation scheme, or a device fingerprint similar (within a margin of error) to that of a legitimate user. Similarly to false rejects, the fuzzy nature of many mimicry-resistant schemes explored in Section~\ref{sec:classi} also introduces the possibility of false accepts, justifying the importance of this new property (M1). Note that false accepts exclude attacks on the integrity of the authentication system (these are covered by other security properties), such as manipulating delay measurements to spoof a location or tampering with client-side code to spoof a device fingerprint. For example, a location-based scheme lacks this benefit if it is susceptible to colocation attacks, \ie where the attacker travels and colocates himself with the user in a highly targeted attack.

\textbf{M2. Resilient-to-Infrequent-Capture-or-Intercept} is a security benefit of schemes in which credentials are not static, but change relatively slowly, \eg personal knowledge questions based on the user's account activity, such as recent transactions. Such non-static credentials, when compromised, limit the duration for which the attacker retains account access. Schemes with horizontal distance from the origin in Fig.~\ref{fig:mimicrychart} (\ie placed in H1, H2, or H3) offer this benefit; see Section~\ref{sec:horizontal} for further discussion of schemes that satisfy this benefit, under the heading \textbf{H1}. 

\textbf{M3. Verifies-Non-Static-Information} is a security benefit of challenge-response based schemes where the server issues a new challenge per authentication; thus an attack that simply captures and uses a static string should not succeed. This benefit is offered by schemes where the server verifies a client's ability to generate a valid response to a challenge, where the response is based on non-static information as opposed to, \eg a static secret. For example, Sound-Proof~\cite{karapanos2015sound} (see Section~\ref{sec:soundproof}) verifies that the ambient sound (which is non-static information) recorded by the user's smartphone matches the ambient sound recorded from the user's access device. OTP schemes may offer this benefit, if (1) the OTPs are not generated based on a static client-stored seed (which would be at risk of being captured by an attacker), and (2) the OTPs expire either upon first use or within a short time frame (\eg 2 minutes), thereby substantially limiting the window for successful replay. Schemes within H2 or H3 in Fig.~\ref{fig:mimicrychart} offer this benefit; see Section~\ref{sec:horizontal} for further discussion of schemes that satisfy this benefit, under the heading \textbf{H2}.

\textbf{M4. Resilient-to-Spoofing} is a security benefit of schemes that leverage measurement techniques (\eg hardware or network-based) that are impractical for an attacker to defeat at scale. For example, CPV \cite{cpvtdsc} is \emph{Resilient-to-Spoofing}, since the measurements cannot be manipulated to make an attacker appear to be in a different location (\ie that of the victim user). Only schemes in H3 of Fig.~\ref{fig:mimicrychart} offer this benefit; see Section~\ref{sec:horizontal} for further discussion of schemes that satisfy this benefit, under the heading \textbf{H3}.

Table~\ref{tab:prims} cells corresponding to benefits M2, M3, and M4 are populated based on the analysis in Section~\ref{sec:placement} (and visualized in Fig.~\ref{fig:mimicrychart}). The following sections evaluate the schemes with respect to the remaining benefits.

\renewcommand\arraystretch{2.4} \setlength\minrowclearance{3.4pt}

\begin{table*}[th!]

\caption{Evaluation of geolocation, device fingerprinting, OTP, and PUF schemes as stand-alone authentication schemes and in combination with passwords. Justification for inclusion: Password authentication is included as a baseline; OTP schemes are widely used in combination with passwords; Sound-Proof~\cite{karapanos2015sound} is a two-factor scheme that also provides some mimicry resistance. \full\ denotes the scheme provides the corresponding benefit (column); \prt\ denotes partial benefit; an empty cell denotes absence of benefit. {\footnotesize*} denotes framework properties introduced herein.}\vspace{10pt}
\label{tab:prims}
\renewcommand{\arraystretch}{1.05}
\scalebox{0.91}{
\begin{tabular}{l|l
|@{}c@{}@{}c@{}@{}c@{}@{}c@{}@{}c@{}@{}c@{}@{}c@{}@{}c@{}@{}c@{}c@{\hskip2pt}
|@{}c@{}@{}c@{}@{}c@{}@{}c@{}@{}c@{}@{}c@{\hskip2pt}
|@{}c@{}@{}c@{}@{}c@{}@{}c@{}@{}c@{}@{}c@{}@{}c@{}@{}c@{}@{}c@{}@{}c@{}@{}c@{}@{}c@{}@{}c@{}@{}c@{}c@{\hskip2pt}
|}

\multicolumn{1}{c}{}&&
\multicolumn{10}{c|}{Usability} &
\multicolumn{6}{c|}{Deployability} &
\multicolumn{15}{c|}{Security} \\[1em]

\rotl{Category} & \multicolumn{1}{c|}{Scheme} &

\rotL{\it U1. Memorywise-Effortless} & 
\rots{\it U2. Scalable-for-Users} & 
\rots{\it U3. Nothing-to-Carry} & 
\rots{\it U4. Physically-Effortless} & 
\rots{\it U5. Easy-to-Learn} & 
\rots{\it U6. Efficient-to-Use} & 
\rots{\it U7. Infrequent-Errors (User errors)} & 
\rots{\it U8. Easy-Recovery-from-Loss} & 
\rots{\it {\footnotesize*}U9. No-False-Rejects (System errors)} & 
\rotR{\it {\footnotesize*}U10. Easy-to-Change-Credentials} &

\rotL{\it D1. Accessible} & 
\rots{\it D2. Negligible-Cost-per-User} & 
\rots{\it D3. Server-Compatible} & 
\rots{\it D4. Browser-Compatible} & 
\rots{\it D5. Mature} & 
\rotR{\it D6. Non-Proprietary} & 

\rotL{\it S1. Resilient-to-Physical-Observation} & 
\rots{\it S2. Resilient-to-Targeted-Impersonation} & 
\rots{\it S3. Resilient-to-Throttled-Guessing} & 
\rots{\it S4. Resilient-to-Unthrottled-Guessing} & 
\rots{\it S5. Resilient-to-Internal-Observation} & 
\rots{\it S6. Resilient-to-Leaks-from-Other-Verifiers} & 
\rots{\it S7. Resilient-to-Phishing} & 
\rots{\it S8. Resilient-to-Physical-Theft} & 
\rots{\it S9. No-Trusted-Third-Party} & 
\rots{\it S10. Requiring-Explicit-Consent} & 
\rots{\it S11. Unlinkable} & 
\rots{\it {\footnotesize*}M1. No-False-Accepts (System errors)} & 
\rots{\it {\footnotesize*}M2. Resilient-to-Infrequent-Capture-or-Intercept} & 
\rots{\it {\footnotesize*}M3. Verifies-Non-Static-Information} & 
\rotR{\it {\footnotesize*}M4. Resilient-to-Spoofing} 
\\ \hline

\multirow{1}{*}{ --}
& P1: Web passwords (PW)
& & &\full & &\full &\full &\prt &\full &\full & \full
&\full &\full &\full &\full &\full &\full	
& &\prt & & & & & &\full &\full &\full &\full & \full & & &	\\\hline

\multirow{4}{*}{\rotatebox[origin=c]{90}{Geolocation}}
& L1: GPS and WPS 
&\full &\full &\full &\full &\full &\full &\prt &\full &\full &
&\full &\full &	&\full &\full &\full	
& & & & & & & &\full & \prt & \prt &\prt &  &  & &	\\

& L2: IP-address based tabulation 
&\full &\full &\full &\full &\full &\full &\prt &\full &  &
&\full &\full &	&\full &\full &\full
& & & & & & & &\full & & &\prt &  &  \full & \prt &	\\

& L3: Measurement-based
&\full &\full &\full &\full &\full &\prt &\prt &\full &  &
&\full &\full &	&\full &\prt &\full	
& & & & & & &  &\full & \full & &\prt &  &  \full & \full &	\\

& L4: Robust location verification 	
&\full &\full &\full &\full &\full &\prt &\prt &\full &  &
&\full &\full &	&\full &	&\full	
& & & & & & & &\full & \full & & \prt &  &  \full & \full & \full	\\

\hline

\multirow{6}{*}{\rotatebox[origin=c]{90}{Device FP}}
& FP1: System parameters/prefs.
&\full &\full &\full &\full &\full &\full &\full & \full  & &
&\full &\full & &\full &\prt & \full 
&\prt &\prt &\full &  &   & &&&\full && \prt &  &  & &\\

& FP2: Audio+Canvas Chal./Resp.
&\full &\full &\full &\full &\full &\full &\full &  \full & &
&\full &\full & &\full &\prt & \full 
&\prt &\prt &\full & & & &&&\full && \prt &   &  \full & \full &	\\

& FP3: HW Sensors
&\full &\full &\full &\full &\full &\full &\full &  \full & &
&\full &\full & &\full & & \full 
&\full & \full & & & & &&&\full && \prt &  &  & &	\\

& FP4: Clock skew
&\full &\full &\full &\full &\full &\full &\full &  \full & &
&\full &\full & &\full & & \full
&\full &\full &  & & & & & & \full & & \prt &  &  \full & \full & \full \\

& FP5: DNS Resolver
&\full &\full &\full &\full &\full &\full &\full &  \full & &
&\full &\full & &\full &\prt & \full
& \prt &\prt & & & & & & & \full & & \prt &  &  \full & \prt &\\

& FP6: Protocol-based
&\full &\full &\full &\full &\full &\full &\full &  \full & &
&\full &\full & &\full &\prt & \full
&\prt &\prt & & & & & & & \full & & \prt &  & \full & & \\\hline

\multirow{4}{*}{\rotatebox[origin=c]{90}{OTP}}
& OTP1: OTP mobile app
&\full &\full & \prt & &\full & & \full & & \full & \full
&\prt & \full  & & \full & \full & \full
& \full & \full  &\full & \full  & &\full &\full & &\full & \full & \full & \full &  & &	\\

& OTP2: OTP USB token
&\full &\full &  & &\full & & \full &  & \full & \full
&\prt &  & & \prt & \full & \full
& \full & \full  &\full & \full  &\full & \full &\full & &\full & \full & \full & \full &   & &	\\

& OTP3: SMS OTP
&\full &\full & \prt & &\full & & \full & \prt & \full & \full
&\prt &  & & \full & \full & \full
& \full & \prt  &\full & \full  & &\full &\full &&\full & \full & & \full &  \full & \prt &	\\

& OTP4: E-Mail OTP
&\full &\full &\full & & \full & &\full & \full & \full & \full
&\prt & \full & & \full &  \full & \full
& \full & \full  & \full & \full  & &\full &\full & \full &\full & \full & \prt & \full &  \full & \prt &	\\

\hline

\multirow{2}{*}{\rotatebox[origin=c]{90}{PUFs}}
& PUF1: Strong PUF
& \full & \full & & \full & \full & \full & \full &  & \full & \full
& \full & &  & & & 
& \full & \full & \full  &\full & \full  & \full & \full &  & \full &  & \full & \full &   & &  	
\\

& PUF2: Weak PUF
& \full & \full & & \full & \full & \full & \full &  & \full & 
& \full & &  & & & \full 
& \full & \full & \full  &\full &   & \full & \full &  &  &  &  & \full & \full & \full &
\\

\hline

\multirow{1}{*}{ --}
& Sound-Proof \cite{karapanos2015sound}
& \full & \prt & \prt  & \full & \full & \full & \full & & \prt & \full
&\full & \full  &  & \full &  &\full 
&\prt &   & \full  &\full &   & \full  & \full &  &\full &  &  \full &  \prt	&  \full & \full &  	
\\

\hline

\multicolumn{33}{c}{\textbf{Combining passwords (PW) with secondary scheme}}\\\hline

\multirow{4}{*}{\rotatebox[origin=c]{90}{Geo.+PW}}
& L1: GPS and WPS 
& & &\full & &\full &\full &\prt &\full & \full &
&\full &\full &	&\full &\full &\full	
& & \prt & & & & & &\full & \prt & \full & \prt  &\full & &  & \\

& L2: IP-address based tabulation 
& & &\full & &\full & \full &\prt &\full &  &
&\full &\full &	&\full &\full &\full
& & \prt & & & & & &\full &  & \full & \prt & \full &  \full & \prt & \\

& L3: Measurement-based
& & &\full & &\full &\prt &\prt &\full &  &
&\full &\full &	&\full &\prt &\full	
& & \prt & & & & & &\full & \full & \full & \prt & \full & \full & \full & \\

& L4: Robust location verification		
& & &\full & &\full &\prt &\prt &\full &  &
&\full &\full &	&\full & &\full	
& & \prt & & & & & &\full & \full & \full & \prt & \full &  \full & \full & \full	\\

\hline

\multirow{6}{*}{\rotatebox[origin=c]{90}{Device FP+PW}}
& FP1: System parameters/prefs.
& & &\full & &\full &\full & \prt & \full & &
&\full &\full & &\full &\prt & \full 
&\prt & \prt &\full &  &   & & & \full &\full & \full & \prt & \full &  & &	\\

& FP2: Audio+Canvas Chal./Resp.
& & &\full & &\full &\full & \prt & \full & &
&\full &\full & &\full &\prt & \full 
&\prt & \prt &\full &  &   & & & \full &\full & \full & \prt & \full &  \full & \full &	\\

& FP3: HW Sensors
& & &\full & &\full &\full & \prt & \full & &
&\full &\full & &\full & & \full 
&\full & \full &  &  &   & & & \full &\full & \full & \prt & \full &  & &	\\

& FP4: Clock skew
& & &\full & &\full &\full & \prt & \full & &
&\full &\full & &\full & & \full
&\full & \full &  &  &   & & & \full &\full & \full & \prt & \full &  \full & \full & \full	\\

& FP5: DNS Resolver
& & &\full & &\full &\full & \prt & \full & &
&\full &\full & &\full &\prt & \full
&\prt & \prt &  &  &   & & & \full &\full & \full & \prt & \full &  \full & \prt &  \\

& FP6: Protocols
& & &\full & &\full &\full & \prt & \full & &
&\full &\full & &\full &\prt & \full
&\prt & \prt &  &  &   & & & \full &\full & \full & \prt & \full & \full &  &  \\

\hline

\multirow{4}{*}{\rotatebox[origin=c]{90}{OTP+PW}}
& OTP1: OTP mobile app
& & & \prt & &\full & & \prt &  & \full & \full
&\prt & \full  & & \full & \full & \full
& \full & \full  &\full & \full  & & \full &\full & \full &\full & \full & \full & \full &  &  &	\\

& OTP2: OTP USB token
& & & & &\full & & \prt &  & \full & \full
&\prt &  & & \prt & \full & \full
& \full & \full  &\full & \full  & \full & \full &\full & \full &\full & \full & \full & \full &  &  &	\\

& OTP3: SMS OTP
& & & \prt & &\full & & \prt & \prt & \full & \full
&\prt &  & & \full & \full & \full
& \full & \prt  &\full & \full  & & \full &\full & \full &\full & \full &  & \full & \full & \prt &	\\

& OTP4: E-Mail OTP
& & & \full & &\full & & \prt & \full & \full & \full
&\prt & \full & & \full &  \full & \full
& \full & \full  &\full & \full  & & \full &\full & \full &\full & \full & \prt  & \full & \full & \prt &	\\

\hline

\multirow{2}{*}{\rotatebox[origin=c]{90}{PUFs}}
& PUF1 + PW
&  &  & &  & \full & \full & \prt &  & \full & \full
&\full & &  & & & 
& \full & \full & \full  &\full & \full & \full & \full &  \full & \full & \full & \full & \full & &   & 	
\\

& PUF2 + PW
&  & & & & \full & \full & \prt &  & \full & 
& \full & &  & &  & \full
& \full & \full & \full  &\full &   & \full & \full & \full  &  & \full  &  & \full &  \full & \full &
\\

\hline

\multirow{1}{*}{ --}
& Sound-Proof + PW \cite{karapanos2015sound}
& & & \prt & & \full & \full & \prt & & \prt	& \full
&\full & \full  &  & \full &  &\full 
&\prt & \prt & \full  &\full &   & \full & \full & \full &\full & \full & \full & \full &  \full & \full &  	
\\

\cline{1-33}

\end{tabular}
}
\end{table*}

\subsection{Impact of Account Recovery on Security}
In practice, authentication schemes are paired with backup mechanisms to help users recover account access if they lose their credentials. The recovery mechanism should not be easier to defeat than the primary scheme, but may sacrifice usability since it is used less frequently.

For conventional password-based authentication, e-mail based password reset is the standard recovery mechanism. Its security relies on the implicit assumption that the user's e-mail account is at least as well-secured as any systems that rely on it for password reset. It is ideally expected that users choose a stronger password for primary e-mail accounts; security-conscious users may also use two-factor authentication. E-mail based recovery can also be used for the geolocation and device fingerprinting schemes evaluated herein; alternatives such as SMS-based OTP are also suitable. For example, in a geolocation-based scheme, a user wishing to login from a new location could be e-mailed a link through which they could confirm their new location.

\subsection{Evaluation of Stand-alone Schemes}
Table~\ref{tab:prims} summarizes our evaluation of four main categories of schemes: geolocation, device fingerprinting, OTPs and PUFs as both stand-alone authentication schemes and in combination with passwords, based on the augmented UDS criteria from Section~\ref{sec:added}. Password authentication is included as reference; OTP schemes are included as they are widely used in combination with passwords. While additional schemes could have been included, those selected were chosen as examples to demonstrate the comparative framework. Each row in the table corresponds to an authentication scheme, and each column to a benefit; a cell with a bullet represents a benefit offered by the scheme, an empty circle represents a benefit partially provided, and an empty cell indicates that the benefit is not provided.

\subsubsection{Web Passwords}
For the new properties, web passwords provide \emph{No-False-Rejects} since a correctly-typed password will never be rejected, \emph{Easy-to-Change-Credentials} since passwords can be easily changed, and \emph{No-False-Accepts} since an exact match is required. However, passwords lack \emph{Resilient-to-Infrequent-Capture-or-Intercept}, \emph{Verifies-Non-Static-Information} and \emph{Resilient-to-Spoofing}, since it is trivial to replay a captured password.

\subsubsection{Location-based Schemes}

Since L1-L4 in Table~\ref{tab:prims} are invisible to the user, they provide most of the usability benefits. L3 (measurement-based) and L4 (location verification) may sometimes take longer than conventionally considered convenient, thus providing only a partial \emph{Efficient-to-Use} benefit. Additionally, all but L1 (GPS/WPS) may miscalculate the location and falsely reject users in some cases, and therefore do not fulfill \emph{No-False-Rejects}. Although L1 (GPS/WPS) may sometimes result in a small error in location calculation, the calculated location will generally remain in the same city, and thus gets a bullet. None of L1-L4 are \emph{Easy-to-Change-Credentials}, since changing credentials would require the legitimate user to change their location.

For deployability, all of L1-L4 are \emph{Accessible}; they do not require any explicit user action. They are \emph{Negligible-Cost-Per-User}, since the infrastructure expense is independent from the number of users being served. They lack \emph{Server-Compatible} as they require server-side changes, but are \emph{Browser-Compatible} (no client-side changes needed). L3 is partially \emph{Mature} since there are indications that it is being used in practice~\cite{nanjee}; L4 is not \emph{Mature}. Finally, \emph{Non-Proprietary} variations for all of L1-L4 are available.

L1-L4 are largely similar to each other in terms of security properties offered, with the exception of \emph{Replay-Resistance} and \emph{Spoofing-Resistance}. Due to the small guessing space, L1-L4 are not \emph{Resilient-to-Throttled-Guessing-Attempts} and \emph{Resilient-to-Unthrottled-Guessing-Attempts}. Since location information must be stored server-side, they are susceptible to the same exposure threats as passwords, namely \emph{Resilient-to-Physical-Observation}, \emph{Resilient-To-Internal-Observation}, \emph{Resilient-to-Leaks-from-Other-Verifiers}, and \emph{Resilient-to-Phishing}. \emph{No-Trusted-Third-Party} is provided by L3-L4 if the website runs their own geolocation infrastructure (\eg deploys their own verifier servers), but not by L2 (IP address tabulation) which typically rely on a third-party service provider, and partially by L1 since GPS does not require a third-party service whereas WPS may. None of L1-L4 are fully \emph{No-False-Accepts}, since other users (both legitimate users and attackers) could be residing near the legitimate user, and thus could be indistinguishable to the server. They are not \emph{Resilient-to-Targeted-Impersonation}, since they are susceptible to targeted colocation attacks. They are not \emph{Requiring-Explicit-Consent}, since they are invisible to the user---except for L1, since a GPS/WPS location request may trigger a browser permission prompt to the user, but if the user allows the browser to remember the preference, no future prompts are displayed. They are all partially \emph{Unlinkable}, since leaking server-stored user location information may narrow the search space for linking together multiple accounts across websites (the attacker may still need to collect additional information).

\subsubsection{Device fingerprinting}
FP1-FP6 deliver most usability benefits, since they are invisible to the user. However, as significant changes in the device configuration (\eg software or hardware upgrade) may substantially change a device fingerprint, they do not provide \emph{No-False-Rejects}. They are \emph{Easy-Recovery-from-Loss} since e-mail based recovery can be used, as with password authentication. They are not \emph{Easy-to-Change-Credentials}, since changing a device fingerprint may require the user to obtain a different device.

For deployability, device fingerprinting schemes are \emph{Accessible}, since they do not require any explicit user action. They are \emph{Negligible-Cost-Per-User}, since the cost of implementation is essentially independent of the number of users. They are not \emph{Server-Compatible}, as server-side implementation is required, but are \emph{Browser-Compatible}. As device fingerprinting has been used for anti-fraud applications \cite{41stpcprint2006, maxminddevic}, but not widely for user authentication, we consider FP1, FP2, FP5, and FP6 partially \emph{Mature}; FP3 and FP4 are not, as they have been demonstrated academically but are not used in practice to our knowledge. FP1-FP6 are generally available via \emph{Non-Proprietary} implementations.

Many of the security properties are shared across FP1-FP6. They lack \emph{No-False-Accepts}, since users that own identical devices (or share the same device) may be indistinguishable from each other (and for techniques such as clockskew, the overall credential space is not large enough to rule out collisions). They lack \emph{Resilient-to-Unthrottled-Guessing-Attempts}, since none of the data collected thus far to our knowledge indicates that the overall distribution of device fingerprints would offer distinguishability of more than about 30 bits \cite{eckersley2010unique,laperdrix2016,bojinovmobile2014,englehardttracking2016,kohnoremote2005,fp2016acsac}. They lack \emph{Resilient-to-Internal-Observation} and \emph{Resilient-to-Phishing}, since an attacker may collect device information by running their own device fingerprinting scripts via XSS attacks or phishing websites.\footnote{Collecting device information helps the attacker, but alone is not enough to defeat mimicry-resistant schemes such as L4.} They lack \emph{Resilient-to-Leaks-from-Other-Verifiers}, since a user's device fingerprint will be similar across websites (varying only based on the particular fingerprinting techniques that each website uses, and the method of storage used), thereby leaking information that can be used to attack users' accounts on different websites. They are each individually partially \emph{Unlinkable} since there may well be multiple users with colliding device fingerprints---the likelihood of this may diminish substantially when combining multiple techniques, however. \emph{Requiring-Explicit-Consent} is not provided, since device fingerprinting is invisible to the user.

The remaining security properties differ across FP1-FP6, as follows. FP3 and FP4 are \emph{Resilient-to-Physical-Observation} and \emph{Resilient-to-Targeted-Impersonation}, since they rely on device-specific manufacturing variations (even across devices of the same model) and therefore can only be determined via measurement; the remaining schemes only partially fulfill these properties since an attacker that visually observes a user's device may obtain or mimic the same device model. FP1 and FP2 have been shown to provide enough distinguishing information to provide \emph{Resilient-to-Throttled-Guessing-Attempts}---to our knowledge the remainder have not, but may collectively provide it if combined together~\cite{fp2016acsac}.

\subsubsection{OTP Schemes}
For usability, OTP1-OTP4 are \emph{Memorywise-Effortless} since they do not require the user to memorize anything; they are \emph{Scalable-for-Users} since they allow the user to set up multiple accounts without impacting usability. OTP4 (e-mail) is \emph{Nothing-to-Carry}, since the user just needs to be able to access their e-mail account, whereas OTP1 (mobile app) and OTP3 (SMS) are partially \emph{Nothing-to-Carry}, assuming users are typically in possession of their mobile phones at all times; OTP2 (USB token) does not fulfill this property. All schemes are \emph{Easy-to-Learn}, but \emph{Inefficient-to-Use} since they require the user to type an extra code. They all provide \emph{Infrequent-Errors} since typos are much less likely with a short 6-digit numerical code, and \emph{No-False-Rejects}. OTP1 and OTP2 lack \emph{Easy-Recovery-From-Loss} since the secret is stored on the device/token; OTP3 partially provides it since there is no secret stored on the phone; OTP4 provides it, since a back-up e-mail address can be used. They all offer \emph{No-False-Rejects}, since a valid OTP will not be rejected. \emph{Easy-to-Change-Credentials} is provided in full by all; a user that needs to switch to a new OTP authenticator could, \eg login with their old authenticator, and then register the new one. 

For deployability, all of OTP1-OTP4 are partially \emph{Accessible}, as blind users would require screen reading software to read the OTP code. Only OTP1 and OTP4 are \emph{Negligible-Cost-per-User}, since SMS incurs a per-message cost to the server,\footnote{Bulk SMS rates can be on the order of a penny each (varying by country). The total cost will vary based on whether the user base is small or in the millions, and on average login frequency.} and USB tokens have a per-user cost. All schemes except for OTP2 (USB tokens) are \emph{Browser-Compatible}; in addition to requiring browser support (at this time, only Google Chrome supports FIDO U2F), OTP2 tokens require a hardware interface (\eg USB-A, USB-C, NFC) that is comaptible with the user's devices---while a variety of hardware tokens are available, they typically only possess one or two interfaces. All of OTP1-OTP4 are \emph{Mature} and with \emph{Non-Proprietary} implementations available.

For security, all of OTP1-OTP4 are \emph{No-False-Accepts}, \emph{Resilient-to-Physical-Observation}, \emph{Resilient-to-Throttled-Guessing}, and \emph{Resilient-to-Unthrottled-Guessing}. OTP1-OTP2 and OTP4 are \emph{Resilient-to-Targeted-Impersonation}, but OTP3 only partially provides this benefit---attackers may transfer the victim's phone number to a new device via a social engineering attack on the users's mobile operator~\cite{smstargetedattack}; the ease of conducting such an attack is subject to the security measures put in place by the mobile operator. They are \emph{Resilient-to-Leaks-from-Other-Verifiers} and \emph{Resilient-to-Phishing}. Only OTP2 (USB tokens) is \emph{Resilient-to-Internal-Observation}---OTP1 is susceptible to theft of the shared seed via malware, and OTP3/OTP4 are susceptible to malware- and network-based~\cite{signalinglink} capture attacks.\footnote{Since the UDS framework combines both network- and device-based eavesdropping/interception into a single property (\emph{Resilient-to-Internal-Observation}), OTP1 and OTP3-OTP4 share the same $y$-coordinate in Fig.~\ref{fig:mimicrychart}, though OTP3 should be lower since it appears to be susceptible to a larger subset of network-based interception attacks~\cite{schneiersms}.} Only the e-mail mechanism (OTP4) is \emph{Resilient-to-Physical-Theft}. They all provide \emph{No-Trusted-Third-Party} and \emph{Requiring-Explicit-User-Consent}. OTP1 and OTP2 are \emph{Unlinkable} since a unique key is used for each website; phone numbers (OTP3) are linkable across sites; e-mail addresses (OTP4) are partially \emph{Unlinkable}, since users may freely create different e-mail aliases. 

\subsubsection{PUFs}
For usability, PUF1 and PUF2 lack \emph{Nothing-to-Carry} and \emph{Easy-Recovery-From-Loss}, since they are tied to the device. PUF1 is \emph{Easy-to-Change-Credentials} since virtually endless challenge/response pairs can be generated; PUF2 is not, since the hardware needs replacement if subjected to a model-building attack. All other usability properties are fulfilled, since no user effort is required.

For deployability, PUF1 and PUF2 are not \emph{Negligible-Cost-per-User}, since hardware would need to be deployed for each user, and they are not \emph{Server-Compatible}, \emph{Browser-Compatible}, or \emph{Mature}. \emph{Non-Proprietary} designs for PUF2 are available~\cite{yu2016pervasive}, but not for PUF1 (to our knowledge).

For security, assuming a large enough space of challenge-response pairs, PUF1 and PUF2 fulfill \emph{No-False-Accepts}, \emph{Resilient-to-Throttled-Guessing}, and \emph{Resilient-to-Unthrottled-Guessing}. PUF1 provides \emph{Resilient-to-Internal-Observation} and \emph{Resilient-to-Phishing} since challenge/response pairs can be exposed without consequence; PUF2 does not provide the former, since it is susceptible to model-building attacks, but it provides the latter since it only responds to challenges originating from a verifier server. Both are \emph{Resilient-to-Leaks-from-Other-Verifiers}. PUF1 and PUF2 are neither \emph{Resilient-to-Physical-Theft} nor \emph{Requiring-Explicit-Consent}. PUF1 has \emph{No-Trusted-Third-Party} and \emph{Unlinkable} since every website can store a separate set of challenge-response pairs, but PUF2 lacks these properties since it typically requires a single server to verify challenge-response pairs (see Section~\ref{sec:pufs}).

\subsubsection{Sound-Proof}
For usability, Sound-Proof provides \emph{Memorywise-Effortless} since there is no secret for the user to memorize. It is partially \emph{Scalable-for-Users}, since only a single device is required, but it appears that under the current architecture each service requires a separate app. It partially provides \emph{Nothing-to-Carry}, since users with a smartphone will typically carry it with them at all times. It is \emph{Easy-to-Learn}, \emph{Efficient-to-Use}, and \emph{Infrequent-Errors} as (aside from installing a smartphone app at setup time) it does not require any user effort to use. It partially provides \emph{No-False-Rejects} and \emph{No-False-Accepts}---the false accept and false reject rates are tunable via a threshold value, and the authors show that the intersection between the two (the equal error rate) is about 0.02\%. It lacks \emph{Easy-Recovery-From-Loss}, as the consequence of losing the phone is similar to that of OTP1. For deployability, it lacks \emph{Server-Compatible} and \emph{Mature}.

For security, Sound-Proof partially provides \emph{Resilient-to-Physical-Observation}, since being able to record sound in the vicinity of the user can break the scheme (it also lacks \emph{Resilient-to-Targeted-Impersonation}, for this reason). It is \emph{Resilient-to-Throttled-Guessing-Attempts} and \emph{Resilient-to-Unthrottled-Guessing-Attempts}. It is not \emph{Resilient-to-Internal-Observation} or \emph{Resilient-to-Theft}, since the scheme can be broken by malware on the phone or by device theft. It is \emph{Resilient-to-Leaks-From-Other-Verifiers} (since there is no static secret to be stored), \emph{Resilient-to-Phishing}, and \emph{No-Trusted-Third-Party}. It is not \emph{Requiring-Explicit-Consent} since it requires no user action. It is \emph{Unlinkable}, since different services can use their own apps on the user's smartphone.

\section{Evaluation of Combined Schemes}
\label{sec:combined}

The evaluation above naturally suggests that schemes offering some degree of mimicry resistance may complement passwords by adding a new security dimension; and invisible schemes such as device fingerprinting and geolocation do so without further burdening users. None of the latter (invisible) schemes, however, seem suitable as a sole mechanism for \uw{} authentication because either the invisibility aspect makes them limited to \dw{} authentication (see Section~\ref{sec:conveyor}), or their security space is so small that they lack the ability to uniquely identify users (\eg suffer from false accepts). This motivates exploring the resultant benefits when these schemes are combined with passwords. This is also useful since some password usability drawbacks might be ameliorated in the medium to long-term by use of browser-based and/or stand-alone password managers. 

When combining two authentication schemes, both sets of credentials should be correct for the user to gain access. This then strengthens security. Ideally, the implementation should limit any partial feedback, which might be beneficial to an attacker using a divide-and-conquer strategy to independently defeat each scheme individually. When password authentication is combined with an invisible scheme, if the correct password is provided but the supporting invisible scheme fails, the server can immediately fall back to a backup scheme such as e-mail OTP. Otherwise, allowing retry authentication attempts benefits attackers, but not legitimate users (except in cases of system error where a retry might succeed, since invisible credentials are not entered by the user). In the event of an attack, this serves to notify the user that their password has been compromised; otherwise, it gives the user an opportunity to reset their supporting authentication mechanism (\eg by registering a new location or new device).

In Table~\ref{tab:prims}, a combined authentication scheme as described above only inherits the intersection of the usability and deployability benefits, and the union of the security benefits---except for \emph{No-Trusted-Third-Party} and \emph{Unlinkable}.

The bottom half of Table~\ref{tab:prims} evaluates the schemes from the top half when used in combination with passwords. The evaluation suggests that either geolocation or device fingerprinting can be combined with passwords to improve security properties, namely \emph{Resilient-to-Physical-Observation}, \emph{Resilient-to-Throttled-Guessing}, \emph{Resilient-to-Infrequent-Capture-or-Intercept}, \emph{Verifies-Non-Static-Information}, and/or \emph{Resilient-to-Spoofing}. To further improve security, geolocation can be combined with one or more device fingerprinting schemes. While combining schemes, select those that maximize the resulting overall benefits, \eg combining L4 with FP2 and/or FP4. As a downside, this would increase the probability of false rejects, assuming the failure of any scheme causes authentication failure. 
PUFs provide many security benefits, suggesting that they could be sufficient as a sole authentication mechanism. However, their major drawback is that they lack most deployability benefits (in the web authentication context).

The usability drawbacks observed for the combined schemes in Table~\ref{tab:prims} are in \emph{Efficient-to-Use} for schemes that introduce a perceptible delay in authenticating the user, \emph{No-False-Rejects} primarily due to variations in measurement or fuzzy matching functions, and \emph{Easy-to-Change}. The deployability drawbacks are in \emph{Server-Compatible} and \emph{Mature}, for schemes that have not yet been widely used in practice.

\section{Concluding Remarks}
\label{sec:conclusion}

While it is not our main objective to provide recommendations---rather the primary focus is in providing the first in-depth treatment of the mimicry-resistance dimension in web authentication schemes and taking this dimension into account in a comparative evaluation framework---we nonetheless give some guidelines which we hope can help advance web authentication research. Our evaluation shows that leveraging schemes with some degree of mimicry-resistance is advantageous to security because they could reduce attack scalability, and can achieve this with minimal impact on usability if the scheme is also invisible. In multi-factor authentication, combining highly usable schemes often enhances the resultant (combined) benefits, even when these offer relatively few security benefits individually since security benefits are often additive (complementary). While multi-factor authentication typically requires all factors to pass in order to grant access, account recovery schemes provide an alternative authentication path that attackers can exploit to bypass the primary authentication mechanism. Therefore, while combining multiple factors strengthens security, the security level of any alternative attack path (\eg account recovery mechanisms) must be correspondingly strengthened.

To the best of our knowledge, this work is the first to systematically investigate the mimicry-resistance aspect of \uw{} and \dw{} authentication schemes. Segmenting the mimicry-resistance dimension into levels of replayability and spoofability helps differentiate the ability of various schemes to resist mimicry attacks. Evaluating schemes using the new dimension, we found that most web authentication schemes have limited to no mimicry resistance as they rely fundamentally on the knowledge of some secret that (once exposed) can be directly used by the attacker for account break-in.

To incorporate mimicry-resistance into an evaluation framework, we add mimicry-resistance properties and explore how mimicry-resistance offers a second dimension orthogonal to exposure-resistance. Plotting these dimensions (Fig.\ref{fig:mimicrychart}) gives a novel representation of attack scalability, where a scheme's ability to resist scalable attacks is conveyed by its distance from the origin. The new framework thus allows for a more comprehensive evaluation of web authentication schemes by their ability to resist both exposure and mimicry, and helps visualize relative resistance to attack scalability, represented as a function of both components. Positioning many representative web authentication schemes on the 2D chart highlights that none explored to date occupy the top-right corner (where schemes highly resistant to scalable attacks would be place).

Using the new framework, we explore mechanisms that offer mimicry-resistant web authentication, and find several such schemes such as some variations of device fingerprinting, several OTP-based schemes, robust Internet geolocation methods, and certain variations of Physically Unclonable Functions (PUFs). Our analysis highlights the degree to which different schemes within the same family of schemes (e.g., different implementations of geolocation) can vary in mimicry-resistance, and our evaluation framework facilitates scheme selection (including from within a family) for maximizing security benefits. For example, in Table 3, some geolocation schemes (e.g., L4) are less susceptible to spoofing than others (e.g., L1). 

A challenging obstacle in reinforcing password authentication with additional schemes is doing so without deployability or usability penalties~\cite{herley2012research}. To evaluate the benefits of combining these four approaches with passwords, we augment the UDS framework with new usability and security properties to reflect mimicry resistance. Using the augmented UDS framework (Table~\ref{tab:prims}) to evaluate the resultant combined schemes revealed that \emph{invisible} techniques---those neither requiring user involvement at set-up nor at login times such as geolocation and device fingerprinting---offer substantial usability advantages and have better deployability than PUFs and OTP-based schemes. 

Our work highlights the advantages of utilizing mimicry-resistant techniques in web authentication, motivating further exploration into invisible and mimicry-resistant techniques, and providing directions on how to explore and evaluate the ability of schemes to resist scalable attacks. 

\section{Acknowledgements}
We thank the anonymous referees for their helpful feedback. The third author acknowledges funding from the Natural Sciences and Engineering Research Council of Canada (NSERC) for both his Canada Research Chair in Authentication and Computer Security, and a Discovery Grant.


\begin{IEEEbiography}[{\includegraphics[width=1in,height=1.25in,clip,keepaspectratio]{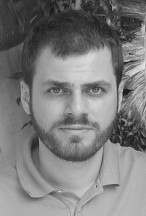}}]{Furkan Alaca} is an Assistant Professor, Teaching Stream at the Department of Mathematical and Computational Sciences at the University of Toronto Mississauga. He completed his PhD (2018) at the School of Computer Science at Carleton University. His research interests include web authentication and device fingerprinting. \end{IEEEbiography}

\begin{IEEEbiography}[{\includegraphics[width=1in,height=1.25in,clip,keepaspectratio]{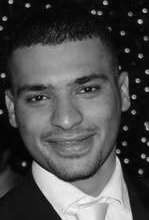}}]{AbdelRahman Abdou} is an Assistant Professor of Computer Science at Carleton University, Canada. In 2018, he was a postdoctoral researcher at the Institute of Information Security, ETH Z\"{u}rich, Switzerland. He received his PhD (2015) in Systems and Computer Engineering from Carleton University. His research interests span SDN security, Internet security (including TLS, DNSSec, authentication, secure BGP, and secure Internet geolocation), and using Internet measurements to understand and solve problems related to Internet systems' security. \end{IEEEbiography} 

\begin{IEEEbiography}[{\includegraphics[width=1in,height=1.25in,clip,keepaspectratio]{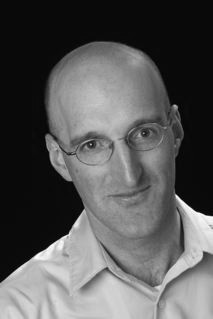}}]{Paul C. van Oorschot} is a Professor of Computer Science at Carleton University, and the Canada Research Chair in Authentication and Computer Security. He was the program chair of USENIX Security 2008 and NDSS 2001-2002; a co-author of the \emph{Handbook of Applied Cryptography}; and a past Associate Editor of \emph{IEEE TDSC}, \emph{IEEE TIFS}, and \emph{ACM TISSEC}. He is a Fellow of the ACM, IEEE, and Royal Society of Canada. His research interests include authentication and Internet security. \end{IEEEbiography}

\vspace{6cm}

\pagebreak
\appendices

\newcommand{\benefit}[1]{\textit{#1}}

\newcommand{\benmem}{
  \benefit{Memorywise-Effortless}}
\newcommand{\benqmem}{
  \quasibenefit{Memorywise-Effortless}}
\newcommand{\bencmem}{U1}

\newcommand{\bensca}{
  \benefit{Scalable-for-Users}}
\newcommand{\benqsca}{
  \quasibenefit{Scalable-for-Users}}
\newcommand{\bencsca}{U2}

\newcommand{\bennottc}{
  \benefit{Nothing-to-Carry}}
\newcommand{\benqnottc}{
  \quasibenefit{Nothing-to-Carry}}
\newcommand{\bencnottc}{U3}

\newcommand{\benman}{
  \benefit{Physically-Effortless}}
\newcommand{\benqman}{
  \quasibenefit{Physically-Effortless}}
\newcommand{\bencman}{U4}

\newcommand{\beneas}{
  \benefit{Easy-to-Learn}}
\newcommand{\benqeas}{
  \quasibenefit{Easy-to-Learn}}
\newcommand{\benceas}{U5}

\newcommand{\beneff}{
  \benefit{Efficient-to-Use}}
\newcommand{\benqeff}{
  \quasibenefit{Efficient-to-Use}}
\newcommand{\benceff}{U6}

\newcommand{\beninf}{
  \benefit{Infrequent-Errors}}
\newcommand{\benqinf}{
  \quasibenefit{Infrequent-Errors}}
\newcommand{\bencinf}{U7}

\newcommand{\benlos}{
  \benefit{Easy-Recovery-from-Loss}}
\newcommand{\benqlos}{
  \quasibenefit{Easy-Recovery-from-Loss}}
\newcommand{\benclos}{U8}

\newcommand{\benacc}{
  \benefit{Accessible}}
\newcommand{\benqacc}{
  \quasibenefit{Accessible}}
\newcommand{\bencacc}{D1}

\newcommand{\benncpu}{
  \benefit{Negligible-Cost-per-User}}
\newcommand{\benqncpu}{
  \quasibenefit{Negligible-Cost-per-User}}
\newcommand{\bencncpu}{D2}

\newcommand{\benser}{
  \benefit{Server-Compatible}}
\newcommand{\benqser}{
  \quasibenefit{Server-Compatible}}
\newcommand{\bencser}{D3}

\newcommand{\benweb}{
  \benefit{Browser-Compatible}}
\newcommand{\benqweb}{
  \quasibenefit{Browser-Compatible}}
\newcommand{\bencweb}{D4}

\newcommand{\benmat}{
  \benefit{Mature}}
\newcommand{\benqmat}{
  \quasibenefit{Mature}}
\newcommand{\bencmat}{D5}

\newcommand{\bennp}{
  \benefit{Non-Proprietary}}
\newcommand{\benqnp}{
  \quasibenefit{Non-Proprietary}}
\newcommand{\bencnp}{D6}

\newcommand{\benrestpo}{
  \benefit{Resilient-to-Physical-Observation}}
\newcommand{\benqrestpo}{
  \quasibenefit{Resilient-to-Physical-Observation}}
\newcommand{\bencrestpo}{S1}

\newcommand{\benrestti}{
  \benefit{Resilient-to-Targeted-Impersonation}}
\newcommand{\benqrestti}{
  \quasibenefit{Resilient-to-Targeted-Impersonation}}
\newcommand{\bencrestti}{S2}

\newcommand{\benrestong}{
  \benefit{Resilient-to-Throttled-Guessing}}
\newcommand{\benqrestong}{
  \quasibenefit{Resilient-to-Throttled-Guessing}}
\newcommand{\bencrestong}{S3}

\newcommand{\benrestofg}{
  \benefit{Resilient-to-Unthrottled-Guessing}}
\newcommand{\benqrestofg}{
  \quasibenefit{Resilient-to-Unthrottled-Guessing}}
\newcommand{\bencrestofg}{S4}

\newcommand{\benrestiox}{
  \benefit{Resilient-to-Internal-Observation}}
\newcommand{\benqrestiox}{
  \quasibenefit{Resilient-to-Internal-Observation}}
\newcommand{\bencrestiox}{S5}

\newcommand{\benlea}{
  \benefit{Resilient-to-Leaks-from-Other-Verifiers}}
\newcommand{\benqlea}{
  \quasibenefit{Resilient-to-Leaks-from-Other-Verifiers}}
\newcommand{\benclea}{S6}

\newcommand{\benphi}{
  \benefit{Resilient-to-Phishing}}
\newcommand{\benqphi}{
  \quasibenefit{Resilient-to-Phishing}}
\newcommand{\bencphi}{S7}

\newcommand{\benthe}{
  \benefit{Resilient-to-Physical-Theft}}
\newcommand{\benqthe}{
  \quasibenefit{Resilient-to-Physical-Theft}}
\newcommand{\bencthe}{S8}

\newcommand{\bennottp}{
  \benefit{No-Trusted-Third-Party}}
\newcommand{\benqnottp}{
  \quasibenefit{No-Trusted-Third-Party}}
\newcommand{\bencnottp}{S9}

\newcommand{\benreq}{
  \benefit{Requiring-Explicit-Consent}}
\newcommand{\benqreq}{
  \quasibenefit{Requiring-Explicit-Consent}}
\newcommand{\bencreq}{S10}

\newcommand{\benunl}{
  \benefit{Unlinkable}}
\newcommand{\benqunl}{
  \quasibenefit{Unlinkable}}
\newcommand{\bencunl}{S11}

\newcommand{\bennfrej}{
  \benefit{No-False-Rejects}}
\newcommand{\bencnfrej}{U9}

\newcommand{\benetoc}{
  \benefit{Easy-to-Change-Credentials}}
\newcommand{\bencetoc}{U10}

\newcommand{\bennfa}{
  \benefit{No-False-Accepts}}
\newcommand{\bencnfa}{M1}

\newcommand{\beninfcap}{
  \benefit{Resilient-to-Infrequent-Capture-or-Intercept}}
\newcommand{\bencinfcap}{M2}

\newcommand{\bennonst}{
  \benefit{Verifies-Non-Static-Information}}
\newcommand{\bencnonst}{M3}

\newcommand{\benresspoof}{
  \benefit{Resilient-to-Spoofing}}
\newcommand{\bencresspoof}{M4}

\section{UDS Properties}
\label{sec:uds}
The definitions of the UDS benefits (properties) can be found in Bonneau et al.~\cite{bonneau2012quest}. Here we list their labels and names for reader convenience. They are grouped in three categories: usabillity (U),  deployability (D), and security (S).

\begin{description}

\item[\bencmem] \benmem

\item[\bencsca] \bensca

\item[\bencnottc] \bennottc

\item[\bencman] \benman

\item[\benceas] \beneas

\item[\benceff] \beneff

\item[\bencinf] \beninf

\item[\benclos] \benlos

\end{description}

\begin{description}

\item[\bencacc] \benacc

\item[\bencncpu] \benncpu

\item[\bencser] \benser

\item[\bencweb] \benweb

\item[\bencmat] \benmat

\item[\bencnp] \bennp

\end{description}

\begin{description}

\item[\bencrestpo] \benrestpo

\item[\bencrestti] \benrestti

\item[\bencrestong] \benrestong

\item[\bencrestofg] \benrestofg

\item[\bencrestiox] \benrestiox

\item[\benclea] \benlea

\item[\bencphi] \benphi

\item[\bencthe] \benthe

\item[\bencnottp] \bennottp

\item[\bencreq] \benreq

\item[\bencunl] \benunl

\end{description}

The new benefits introduced herein are listed below. The mimicry-resistance (M) benefits are a new sub-category of security benefits. For definitions of these new properties, see Section \ref{sec:added}. 

\begin{description}
\item[\bencnfrej] \bennfrej
\item[\bencetoc] \benetoc
\item[\bencnfa] \bennfa
\item[\bencinfcap] \beninfcap
\item[\bencnonst] \bennonst
\item[\bencresspoof] \benresspoof
\end{description}

 \balance

\end{document}